\documentclass[11pt]{article}
\usepackage{amsfonts}
\usepackage{latexsym,amsmath}
\usepackage{amssymb,array}
\usepackage{calc}
\RequirePackage[dvips]{graphicx}
\usepackage{graphics}
\usepackage[colorlinks=true]{hyperref}
\hypersetup{urlcolor=blue, citecolor=red}
\makeatletter \oddsidemargin 0in \evensidemargin 0in \textwidth
16cm\textheight 20cm 
\begin{document}
\newcommand{\mf}{\mathbf}
\newcommand{\ov}{\overline}
\newcommand{\om}{\omega}
\newcommand{\ga}{\gamma}
\newcommand{\cd}{\circledast}
\newcommand{\eq}{\Leftrightarrow}
\newtheorem{t1}{Theorem}
\newtheorem{pro}{Proposition}
\newtheorem{remark}{Remark}
\newtheorem{ce}{Counterexample}
\newtheorem{coro}{Corollary}
\newtheorem{d1}{Definition}
\newtheorem{n1}{Notation}
\newtheorem{example}{Example}
\newtheorem{l1}{Lemma}
\title{\bf Ordering properties of the smallest and largest lifetimes in Gompertz-Makeham model} 
\author{Amarjit Kundu\\Department of
Mathematics\\
Raigunj University\\West Bengal, India. \and Shovan Chowdhury\footnote{Corresponding
author; e-mail: meetshovan@gmail.com, shovanc@iimk.ac.in}\\Quantitative Methods and Operations Management Area\\Indian Institute of Management, Kozhikode\\Kerala, India. \and Narayanaswamy Balakrishnan\\Department of Mathematics and Statistics\\ McMaster University\\ Hamilton, Canada
}
\maketitle
{\bf Abstract}:
The Gompertz-Makeham distribution, which is used commonly to represent lifetimes based on laws of mortality, is one of the most popular choices for mortality modelling in the field of actuarial science. This paper investigates ordering properties of the smallest and largest lifetimes arising from two sets of heterogeneous groups of insurees following respective Gompertz-Makeham distributions. Some sufficient conditions are provided in the sense of usual stochastic ordering to compare the smallest and largest lifetimes from two sets of dependent variables. Comparison results on the smallest lifetimes in the sense of hazard rate ordering and ageing faster ordering are established for two groups of heterogeneous independent lifetimes. Under similar set-up, no reversed hazard rate ordering is shown to exist between the largest lifetimes with the use of a counter-example. Finally, we present sufficient conditions to stochastically compare two sets of independent heterogeneous lifetimes under random shocks by means of usual stochastic ordering. Such comparisons for the smallest lifetimes are also carried out in terms of hazard rate ordering. 


\section{Introduction}
\setcounter{equation}{0}
\hspace*{0.3 in} Life insurance companies offer a wide range of life insurance and annuity products to its customers. The companies bear a huge liability for selling such products in terms of buying risk of death of an insured. Eventually, such liabilities necessitate the companies to gain knowledge on the mortality patterns of a community in a place. Modeling future mortality helps assessing the reserve and capital required for a portfolio of immediate or deferred annuities held by the life insurance company. It is also necessary to model future mortality for better pricing of the products and to estimate the claim amount from a portfolio sold by life insurance contracts.
\\\hspace*{0.3 in} The Gompertz-Makeham distribution, which represents survival time based on laws of mortality, is one of the popular choices for mortality modelling in actuarial science. In 1825, Gompertz~(\cite{go}) presented his version of the survival model for human mortality, based on the notion that human mortality increases exponentially with age. The hazard function of the Gompertz model is given by 
\begin{equation}\label{l1}
h(x)=\alpha e^{\beta x},~x>0,~\alpha>0,~\beta>0,
\end{equation}
where the parameters $\alpha$ and $\beta$ can be interpreted, respectively, as the initial mortality and the exponential coefficient of mortality increase. The model proposed by Gompertz is one of the most influential parametric mortality model in the literature. About three decades later, in 1860, Makeham~(\cite{ma}) noticed that Gompertz's model was not adequate for higher ages and improved the model by adding an age-independent constant $\lambda>0$ to the exponential growth, which captures the risk of death from all causes that do not depend on age. This model is known as Gompertz-Makeham (GM) model with hazard function given by 
\begin{equation}\label{l2}
h(x)=\alpha e^{\beta x}+\lambda,~x>0,~\alpha>0,~\beta>0,~\lambda>0.
\end{equation}
The hazard function in (\ref{l2}) leads to the survival function of the GM model as  
\begin{equation}\label{l3}
\bar{F}(x)=e^{-\lambda x-\frac{\alpha}{\beta}\left(e^{\beta x}-1\right)},~x>0,~\alpha,\beta,\lambda>0.
\end{equation}
The random variable (rv) $X$ represents life length or survival time of the GM distribution, and will be denoted by GM($\alpha,\beta,\lambda$) hereafter. It may also be noted that there exists different parametrizations of the GM distribution in the literature (see Johnson et al.~(\cite{jo})). While GM model is investigated closely with respect to its interesting properties and applications in insurance and actuarial science (see Melnikov and Romaniuk~\cite{mel}, Richards~\cite{ri}, Lockwood~\cite{lo}, Andreopoulos et al.\cite{an}), ordering properties of order statistics from this distribution under heterogeneous set-up have not been studied so far. This is the primary motivation for the work carried out here.\\
\hspace*{0.3 in} Stochastic ordering has been a topic of great interest in various areas including management science, financial economics, insurance, actuarial science, operations research, reliability theory and survival analysis. If $X_{1:n}\leq X_{2:n}\leq\ldots\leq X_{n:n}$ denote the order statistics arising from the random variables $X_1, X_2,\ldots,X_n$, then the minimum and maximum lifetimes correspond to $X_{1:n}$ and $X_{n:n},$ respectively. Much work has been carried out in the past two decades to compare lifetimes of the smallest and largest order statistics from heterogeneous independent rvs  with certain underlying distributions. One may refer to Dykstra et al.~\cite{dkr11}, Zhao and Balakrishnan~(\cite{zb11.2}), Balakrishnan et al.~\cite{ba1}, Torrado and Kochar~\cite{tr11}, Fang and Balakrishnan~\cite{fb}, Kundu et al.~\cite{kun1}, Kundu and Chowdhury~(\cite{kun2},\cite{kun3}), Chowdhury and Kundu~\cite{ch} and the references therein for more details. Such comparisons are generally carried out with the assumption that the units of the sample die with certainty. In practice, the units may experience random shocks which eventually doesn't guarantee its death. Stochastic comparison of two systems under random shocks have been discussed recently by Barmalzan and Najafabadi~\cite{ba5}, Barmalzan et al.~\cite{bar3}, and Balakrishnan et al.~\cite{ba2} in the context of insurance claims. However, in many practical situations, the units of a sample may have a structural dependence which would result in a set of statistically dependent variables. The dependence structure of the components have been investigated by Navarro and Spizzichino~\cite{ne}, Rezapour and Alamatsaz~\cite{re}, and Li and Fang~\cite{li1} with the use of copulas. \\ 
 \hspace*{0.3 in} As mentionmed earlier, stochastic comparison of the smallest and largest claim amounts or aggregate claims have been carried out so far in the context of insurance and actuarial science. Mortality conditions of insurees are closely associated with the amount claimed by them. GM distribution is a popular choice for describing human mortality and establishing actuarial tables. Thus, we are motivated to compare the maximum or the minimum lifetimes of two groups of insurees stochastically under three different situations. In this sense, the paper distinguishes itself from other work mentioned above. In the first situation, the lives of the individuals in each group are assumed to be independent of each other. For example, we may be interested to compare minimum and maximum lifetimes of two groups of insurees from two different geographical regions. The second situation assumes that the individuals in each group are dependent on each other. Such a situation arises when we compare minimum and maximum lifetimes of the insurees from two households in the same geographical region. In the third situation, minimum and maximum of the lifetimes of two groups of insurees are compared under random shock when the lifetimes are independently distributed. Under this set up, lifetimes of two groups of insurees may be compared with two different occupational hazards. \\The rest of this paper is organized as follows. In Section 2, we have given the required definitions and some useful lemmas that are used throughout the paper. Results concerning stochastic comparison of the minimum (maximum) lifetimes of two samples of insurees with heterogeneous independent and dependent lifetimes are derived in Section~3~(Section~4). Section~5 discusses some results concerning the comparison of the minimum/maximum lifetimes of two samples of insurees with heterogeneous independent lifetimes under random shock. 

\section{Preliminaries}
\setcounter{equation}{0}
\hspace*{0.3 in} For two absolutely continuous random variables $X$ and $Y$ with distribution functions $F\left(\cdot\right)$ and $G\left(\cdot\right)$, survival functions $\overline F\left(\cdot\right)$ and $\overline G\left(\cdot\right)$, density functions $f\left(\cdot\right)$ and $g\left(\cdot\right)$, hazard rate functions $r\left(\cdot\right)$ and $s\left(\cdot\right)$ and reversed hazard rate functions ${\tilde r(\cdot)}$ and ${\tilde s(\cdot)},$ respectively, $X$ is said to be smaller than $Y$ in $(i)$ {\it likelihood ratio order} (denoted by $X\leq_{lr}Y$), if, for all $t$, $\frac{g(t)}{f(t)}$ increases in $t$, $(ii)$ {\it hazard rate order} (denoted by $X\leq_{hr}Y$), if, for all $t$, $\frac{\overline G(t)}{\overline F(t)}$ increases in $t$ or equivalently $r(t)\ge s(t)$, 
$(iii)$ {\it reversed hazard rate order} (denoted by $X\leq_{rh}Y$), if, for all $t$, $\frac{G(t)}{ F(t)}$ increases in $t$ or equivalently ${\tilde r(t)}\leq {\tilde s(t)}$, 
$(iv)$ {\it Ageing faster order in terms of the hazard rate order} (denoted by $X\leq_{R-hr}Y$), if, for all $t$, $\frac{r_X(t)}{r_Y(t)}$ increases in $t$, and $(v)$  {\it usual stochastic order} (denoted by $X\leq_{st}Y$), if $F(t)\ge G(t)$ for all $t$. One may refer to Sengupta and Deshpande~\cite{se}, and Li and Li~\cite{lili} for more details on relative ageing, also known as ageing faster ordering in terms of hazard rate or reversed hazard rate ordering. For more elaborate details on stochastic orders, see Shaked and Shanthikumar \cite{shak1}. 


Let $x_{(1)}\le x_{(2)}\le\cdots\le x_{(n)}$ be the
increasing arrangement and $x_{[1]}\ge x_{[2]}\ge\cdots\ge x_{[n]}$ be the decreasing arrangement of the components of the vector $\mathbf{x} $=$\left(x_1,\;x_2,\ldots, x_n\right)$. 
The following definitions can be found in Marshall \emph{et al.} \cite{Maol}.\\
\hspace*{0.3 in}Let $I^n$ denote an $n$-dimensional Euclidean space, where $I\subseteq\mathbb{R}$. Further, let $\mathbf{x}=(x_1,x_2,\dots,x_n)\in I^n$ and $\mathbf{y}=(y_1,y_2,\dots,y_n)\in I^n$ be any two vectors.
\begin{d1}
\begin{enumerate}
\item[(i)] The vector $\mathbf{x} $ is said to majorize vector $\mathbf{y} $ (written as $\mathbf{x}\stackrel{m}{\succeq}\mathbf{y}$) if
\begin{equation*}\sum_{i=1}^j x_{[i]}\ge\sum_{i=1}^j y_{[i]},\;j=1,\;2,\;\ldots, n-1,\;\;and\; \;\sum_{i=1}^n x_{[i]}=\sum_{i=1}^n y_{[i]},\end{equation*}
or equivalently,
\begin{equation*}
\sum_{i=1}^j x_{(i)}\le\sum_{i=1}^j y_{(i)},\;j=1,\;2,\;\ldots, n-1,\;\;and \;\;\sum_{i=1}^n x_{(i)}=\sum_{i=1}^n y_{(i)};
\end{equation*} 
\item [(ii)] The vector $\mathbf{x}$ is said to weakly supermajorize the vector $\mathbf{y}$
 (written as $\mathbf{x}\stackrel{\rm w}{\succeq} \mathbf{y}$) if
 \begin{eqnarray*}
  \sum\limits_{i=1}^j x_{(i)}\leq \sum\limits_{i=1}^j y_{(i)}\quad \text{for}\;j=1,2,\dots,n;
 \end{eqnarray*}
 \item [(iii)] The vector $\mathbf{x}$ is said to weakly submajorize the vector $\mathbf{y}$
 (written as $\mathbf{x}\;{\succeq}_{\rm w} \;\mathbf{y}$) if
 \begin{eqnarray*}
\sum\limits_{i=j}^n x_{(i)}\geq \sum\limits_{i=j}^n y_{(i)}\quad \text{for}\;j=1,2,\dots,n.
 \end{eqnarray*}
\end{enumerate}
\end{d1}

\begin{d1}\label{de2}
A function $\psi:\mathbb{R}^n\rightarrow\mathbb{R}$ is said to be Schur-convex (resp. Schur-concave) on $\mathbb{R}^n$ if 
\begin{equation*}
\mathbf{x}\stackrel{m}{\succeq}\mathbf{y} \;\text{implies}\;\psi\left(\mathbf{x}\right)\ge (\text{resp. }\le)\;\psi\left(\mathbf{y}\right)\;for\;all\;\mathbf{x},\;\mathbf{y}\in \mathbb{R}^n.
\end{equation*}
\end{d1}
\begin{n1}
Let us introduce the following notations:
\begin{enumerate}
\item[(i)] $\mathcal{D}_{+}=\left\{\left(x_{1},x_2,\ldots,x_{n}\right):x_{1}\ge x_2\ge\ldots\ge x_{n}> 0\right\}$,
\item[(ii)] $\mathcal{E}_{+}=\left\{\left(x_{1},x_2,\ldots,x_{n}\right):0< x_{1}\leq x_2\leq\ldots\leq x_{n}\right\}$.
\end{enumerate}
\end{n1}
We now introduce the following lemmas which will be used in subsequent sections to establish our main results. 
 \begin{l1}\label{l6}
 $\left(\text{Lemma 3.1 of Kundu \emph{et al.}~\cite{kun1}}\right)$~Let $\varphi:\mathcal{D_{+}}\rightarrow \mathbb{R}$ be a function, continuously differentiable on the interior of $\mathcal{D_{+}}$. Then, for $\mf{x},\mf{y}\in \mathcal{D_{+}}$,
 \begin{eqnarray*}
  \mf{x}\stackrel{m}\succeq\mf{y}\;\text{implies}\;\varphi(\mf{x})\ge (\text{resp.}\;\leq)\;\varphi(\mf{y})
 \end{eqnarray*}
if, and only if,
$$\varphi_{(k)}(\mf{z})\;\text{is decreasing (resp. increasing) in}\;k=1,2,\dots,n,$$
where $\varphi_{(k)}(\mf{z})=\partial\varphi(\mf{z})/\partial z_k$ denotes the partial derivative of $\varphi$ with respect to its $k$th argument.
\end{l1}
\begin{l1}\label{l7}
 $\left(\text{Lemma 3.3 of Kundu \emph{et al.}~\cite{kun1}}\right)\;$~Let $\varphi:\mathcal{E_{+}}\rightarrow \mathbb{R}$ be a function, continuously differentiable on the interior of $\mathcal{E_{+}}$.
 Then, for $\mf{x},\mf{y}\in \mathcal{E_{+}}$,
 \begin{eqnarray*}
  \mf{x}\stackrel{m}\succeq\mf{y}\;\text{implies}\;\varphi(\mf{x})\ge (\text{resp.}\;\leq)\;\varphi(\mf{y})
 \end{eqnarray*}
if, and only if,
$$\varphi_{(k)}(\mf{z})\;\text{is increasing (resp. decreasing) in}\;k=1,2,\dots,n,$$
where $\varphi_{(k)}(\mf{z})=\partial\varphi(\mf{z})/\partial z_k$ denotes the partial derivative of $\varphi$ with respect to its $k$th argument.
\end{l1}
\begin{l1}\label{l33}$\left(\text{Lemma 2.2 of Kundu and Chowdhury~\cite{kun2}}\right)$~Let $\varphi: I^n\rightarrow \mathbb{R}$. 
Then $$(a_1,a_2,\ldots,a_n){\succeq} _{w} (b_1,b_2,\ldots,b_n)\; \text{implies} \;\varphi(a_1,a_2,\ldots,a_n)\geq (\text{resp. }\leq)\; \varphi(b_1,b_2,\dots,b_n)$$ 
if, and only if, $\varphi$ is increasing (resp. decreasing) and Schur-convex (resp. Schur-concave) on $I^n$. Similarly,
 $$(a_1,a_2,\ldots,a_n)\stackrel{w}\succeq (b_1,b_2,\ldots,b_n)\; \text{implies} \;\varphi(a_1,a_2,\dots,a_n)\geq (\text{resp. }\leq)\;\varphi(b_1,b_2,\dots,b_n)$$ 
if, and only if, $\varphi$ is decreasing (resp. increasing) and Schur-convex (resp. Schur-concave) on $I^n$.
 \end {l1}
\hspace*{0.3 in} Now, let us recall that a copula associated with a multivariate distribution function $F$ is a function $C:\left[0,1\right]^n\longmapsto\left[0,1\right]$ such that $F(x)=C\left(F_{1}(X_1),..., F_{n}(X_n)\right),$ where the $F_i$'s, $1\leq i\leq n,$ are the univariate marginal distribution functions of $X_i$'s. Similarly, a survival copula associated with a multivariate survival function $\overline{F}$ is a function $\overline{C}:\left[0,1\right]^n\longmapsto\left[0,1\right]$ such that
$$\overline{F}(x)=P\left(X_1>x_1,...,X_n>x_n\right)=\overline{C}\left(\overline{F}_1(x_1),...,\overline{F}_n(x_n)\right),$$ 
where, for $1\leq i\leq n$, $\overline{F}_i(\cdot)=1-F_i(\cdot)$  are the univariate survival functions. Specifically, a copula $C$ is Archimedean if there exists a generator $\psi:\left[0,\infty\right]\longmapsto\left[0,1\right]$ such that
$$C\left(\mathbf{u}\right)=\psi\left(\psi^{-1}(u_1),...,\psi^{-1}(u_d)\right).$$
For $C$ to be an Archimedean copula, it is sufficient and necessary that $\psi$ satisfies $(i)$ $\psi(0)=1$ and $\psi(\infty)=0$ and $(ii)$ $\psi$ is $d-$monotone, i.e., $\frac{(-1)^k d^k \psi(s)}{ds^k}\ge 0$ for $k\in \left\{0,1,...,d-2\right\}$ and $\frac{(-1)^{d-2} d^{d-2} \psi(s)}{ds^{d-2}}$ is decreasing and convex. Archimedean copulas
cover a wide range of dependence structures including the independence copula and the Clayton copula. For more details on Archimedean copula, one must refer to Nelsen~\cite{ne} and McNeil and N$\check{e}$slehov$\acute{a}$~\cite{mc}. In this paper, Archimedean copula is specifically used to model the dependence structure among random variables in a sample. The following important lemma is used in the subsequent sections to prove some of the important results. 
\begin{l1}\label{l11}
 $\left(\text{Li and Fang~\cite{li1}}\right)\;$~For two n-dimensional Archimedean copulas $C_{\psi_1}\left(\mathbf{u}\right)$ and $C_{\psi_2}\left(\mathbf{u}\right)$, with $\phi_2=\psi_{2}^{-1}=sup\left\{x\in\mathbb{R}:\psi(x)>u\right\}$ as the right continuous inverse, if $\phi_2\circ\psi_1$ is super-additive, then $C_{\psi_1}\left(\mathbf{u}\right)\leq C_{\psi_2}\left(\mathbf{u}\right)$ for all $\mathbf{u}\in [0,1]^n.$ Recall that a function $f$ is said to be super-additive if $f(x+y)\ge f(x) + f(y)$ for all $x$ and $y$ in the domain of $f$.
 \end{l1} 

\section{Comparison of two sample minima}
\setcounter{equation}{0}
\hspace*{0.3 in} In this section, some results on stochastic comparison of smallest lifetimes of two samples are derived when the lifetimes follow GM distributions. The comparison is carried out under two scenarios: first when the underlying variables have a dependent structure sharing Archimedean copulas and second when the underlying variables are independently distributed. \\
\subsection{\small Dependent case}
\hspace*{0.3 in} Let, $X$ and $Y$ be two random variables having distribution functions $F(\cdot)$ and $G(\cdot)$ respectively. Let $X_i\sim GM\left(\alpha_i,\beta_i,\lambda_i\right)$ and $Y_i\sim GM\left(\alpha_i^{*},\beta_i^{*},\lambda_i^{*}\right)$ ($i=1,2,\ldots,n$) be two sets of $n$ dependent random variables with Archimedean copulas having generators $\psi_1~\left(\phi_1=\psi_{1}^{-1}\right)$ and $\psi_2~\left(\phi_2=\psi_{2}^{-1}\right),$ respectively. Further, let $\overline{G}_{1:n}\left(\cdot\right)$ and $\overline{H}_{1:n}\left(\cdot\right)$ be the survival functions of $X_{1:n}$ and $Y_{1:n},$ respectively. Then, 
\begin{equation*}
\overline{G}_{1:n}\left(x\right)=\psi_1\left[\sum_{k=1}^n \phi_1\left(e^{-\lambda_k x-\frac{\alpha_k}{\beta_k}\left(e^{\beta_k x}-1\right)}\right)\right],~x>0,
\end{equation*}
and
\begin{equation*}
\overline{H}_{1:n}\left(x\right)=\psi_2\left[\sum_{k=1}^n \phi_2\left(e^{-\lambda_k^{*} x-\frac{\alpha_k^{*}}{\beta_k^{*}}\left(e^{\beta_k^{*} x}-1\right)}\right)\right],~x>0.
\end{equation*}
\hspace*{0.3 in} Let $r_X(u)$ and $r_Y(u)$ be the hazard rate functions of the random variables $X$ and $Y,$ respectively. The first result shows that usual stochastic ordering exists between $X_{1:n}$ and $Y_{1:n}$ under majorization order of the initial mortality parameters.
\begin{t1}\label{th1}
Let $X_1,X_2,...,X_n$ be a set of dependent random variables sharing Archimedean copula having generator $\psi_1,$ such that $X_i\sim GM\left(\alpha_i,\beta_i,\lambda_i\right),~i=1,2,...,n$. Let $Y_1,Y_2,...,Y_n$ be another set of dependent random variables sharing Archimedean copula having generator $\psi_2,$ such that $Y_i\sim GM\left(\alpha_i^{*},\beta_i,\lambda_i\right) ,~i=1,2,...,n$. Assume that $\mbox{\boldmath $\alpha$}, \mbox{\boldmath $\alpha^*$}, \mbox{\boldmath $\beta$}\;and\; \mbox{\boldmath $\lambda$}\in \mathcal{D}_+$ (or $\mathcal{E}_+$). Further, suppose $\phi_2\circ\psi_1$ is super-additive, $\psi_1$ or $\psi_2$ is log-convex. Then, $\mbox{\boldmath $\alpha$}\succeq_w \mbox{\boldmath $\alpha^*$}$ implies $X_{1:n}\leq_{st}Y_{1:n}$.
\end{t1}
{\bf Proof:} By Lemma \ref{l11}, the super-additivity of $\phi_2\circ\psi_1$ implies that 
\begin{equation}\label{e1}
\psi_1\left[\sum_{k=1}^n \phi_1\left(e^{-\lambda_k x-\frac{\alpha_k}{\beta_k}\left(e^{\beta_k x}-1\right)}\right)\right]\leq \psi_2\left[\sum_{k=1}^n \phi_2\left(e^{-\lambda_k x-\frac{\alpha_k}{\beta_k}\left(e^{\beta_k x}-1\right)}\right)\right].
\end{equation}
Let us assume that
$$\psi_2\left[\sum_{k=1}^n \phi_2\left(e^{-\lambda_k x-\frac{\alpha_k}{\beta_k}\left(e^{\beta_k x}-1\right)}\right)\right]=\Psi(\mbox{\boldmath $\alpha$}).$$ 
Upon differentiating $\Psi(\mbox{\boldmath $\alpha$})$ with respect to $\alpha_i$, we get
\begin{equation}\label{e002}
\begin{split}
\frac{\partial \Psi}{\partial \alpha_i}&=-\psi_{2}^{'}\left[\sum_{k=1}^n \phi_2\left(e^{-\lambda_k x-\frac{\alpha_k}{\beta_k}\left(e^{\beta_k x}-1\right)}\right)\right]\\&\quad\frac{\psi_{2}\left[\phi_2\left(e^{-\lambda_i x-\frac{\alpha_i}{\beta_i}\left(e^{\beta_i x}-1\right)}\right)\right]}{\psi_{2}^{'}\left[\phi_2\left(e^{-\lambda_i x-\frac{\alpha_i}{\beta_i}\left(e^{\beta_i x}-1\right)}\right)\right]}\frac{e^{\beta_i x}-1}{\beta_i}\leq 0,
\end{split}
\end{equation}
yielding $\Psi(\mbox{\boldmath $\alpha$})$ to be decreasing in each $\alpha_i$. It can be easily verified that $\frac{e^x-1}{x}$ is an increasing function in $x.$ So, for $i\leq j,$ if $\beta_i\geq (\leq)\beta_j,$ then 
\begin{equation}\label{e003}
x\frac{e^{\beta_i x}-1}{x \beta_i}\geq (\leq)x\frac{e^{\beta_j x}-1}{x \beta_j}.
\end{equation}
Again, it can be easily shown that $\left(e^{-\lambda_i x-\frac{\alpha_i}{\beta_i}\left(e^{\beta_i x}-1\right)}\right)$ is a decreasing function in $\alpha_i,$ $\beta_i,$ and $\lambda_i.$ Moreover, as $\psi_2$ is log-convex, implying that $\frac{\psi_{2}(u)}{\psi_{2}^{'}(u)}$ is decreasing in $u,$ then, for all $i\leq j$, $\alpha_i\geq \alpha_j,$ $\beta_i\geq \beta_j,$ and $\lambda_i\geq \lambda_j,$ it can be shown that
\begin{equation}\label{e004}
 \frac{\psi_2\left(u_i\right)}{\psi_{2}^{'}\left(u_i\right)}\leq (\geq)\frac{\psi_2\left(u_j\right)}{\psi_{2}^{'}\left(u_j\right)},  
\end{equation} 
where $u_i=\phi_2\left(e^{-\lambda_i x-\frac{\alpha_i}{\beta_i}\left(e^{\beta_i x}-1\right)}\right).$ \\
Using (\ref{e003}) and (\ref{e004}) in (\ref{e002}), it can be easily shown that $\frac{\partial \Psi}{\partial \alpha_i}-\frac{\partial \Psi}{\partial \alpha_j}\leq (\ge) 0$. Thus, by Lemma \ref{l6} (Lemma \ref{l7}), we have $\Psi(\mbox{\boldmath $\alpha$})$ to be s-concave in $\mbox{\boldmath $\alpha$}$. Thus using Lemma \ref{l33} and \ref{e1}, the result is proved. \hfill$\Box$\\

The next two results discuss stochastic ordering between $X_{1:n}$ and $Y_{1:n}$ under majorization order of the coefficient of mortality parameters and age independent constant parameters separately when $\psi_1$ or $\psi_2$ is log-convex. As these results can be proved in a way similar to Theorem \ref{th1}, we do not present the proofs here for the sake of brevity. 
\begin{t1}\label{th2}
Let $X_1,X_2,...,X_n$ be a set of dependent random variables sharing Archimedean copula having generator $\psi_1,$ such that $X_i\sim GM\left(\alpha_i,\beta_i,\lambda_i\right),~i=1,2,...,n$. Let $Y_1,Y_2,...,Y_n$ be another set of dependent random variables sharing Archimedean copula having generator $\psi_2,$ such that $Y_i\sim GM\left(\alpha_i,\beta_i^{*},\lambda_i\right) ,~i=1,2,...,n$. Assume that $\mbox{\boldmath $\alpha$},\mbox{\boldmath $\beta$},\mbox{\boldmath $\beta^*$}\;and\; \mbox{\boldmath $\lambda$}\in \mathcal{D}_+$ (or $\mathcal{E}_+$). Further, suppose $\phi_2\circ\psi_1$ is super-additive, $\psi_1$ or $\psi_2$ is log-convex. Then, $\mbox{\boldmath $\beta$}\succeq_w \mbox{\boldmath $\beta^*$}$ implies $X_{1:n}\leq_{st}Y_{1:n}$.
\end{t1}
 
\begin{t1}\label{th3}
Let $X_1,X_2,...,X_n$ be a set of dependent random variables sharing Archimedean copula having generator $\psi_1,$ such that $X_i\sim GM\left(\alpha_i,\beta_i,\lambda_i\right),~i=1,2,...,n$. Let $Y_1,Y_2,...,Y_n$ be another set of dependent random variables sharing Archimedean copula having generator $\psi_2,$ such that $Y_i\sim GM\left(\alpha_i,\beta_i,\lambda_i^{*}\right),~i=1,2,...,n$. Assume that $\mbox{\boldmath $\alpha$},\mbox{\boldmath $\beta$},\mbox{\boldmath $\lambda$}, \mbox{\boldmath $\lambda^*$}\in \mathcal{D}_+$ (or $\mathcal{E}_+$). Further, suppose $\phi_2\circ\psi_1$ is super-additive, $\psi_1$ or $\psi_2$ is log-convex. Then, $\mbox{\boldmath $\lambda$}\succeq_w \mbox{\boldmath $\lambda^*$}$ implies $X_{1:n}\leq_{st}Y_{1:n}$.
\end{t1}

It is important to point out that Theorem \ref{th1} guarantees that the smallest lifetime of the first sample of insurees is stochastically smaller than that of the second sample of insurees with common heterogeneous exponential coefficient of mortality parameter and age-independent parameter, when the initial mortality of the first sample of insurees majorizes that of the second sample. Theorems \ref{th2} and \ref{th3} can be interpreted in a similar manner as well!
\subsection{\small Heterogenous independent case}
\hspace*{0.3 in} For $i=1,2,\ldots,n$, let $X_i$ and $Y_i$ be two sets of $n$ independent random variables following $GM$ distribution with parameters ($\alpha_i,\beta_i,\lambda_i$) and ($\alpha_i^*,\beta_i^*,\lambda_i^*$) respectively.
If $\overline{G}_{1:n}\left(\cdot\right)$ and $\overline{H}_{1:n}\left(\cdot\right)$ are the survival functions of $X_{1:n}$ and $Y_{1:n},$ respectively, then evidently
\begin{equation}\label{11}
\overline{G}_{1:n}\left(x\right)=\prod_{k=1}^n \left[e^{-\lambda_k x-\frac{\alpha_k}{\beta_k}\left(e^{\beta_k x}-1\right)}\right]
\end{equation}
and
\begin{equation}\label{12}
\overline{H}_{1:n}\left(x\right)=\prod_{k=1}^n \left[e^{-\lambda_k^{*} x-\frac{\alpha_k^{*}}{\beta_k^{*}}\left(e^{\beta_k^{*} x}-1\right)}\right].
\end{equation}
Then, if $r_{1:n}(\cdot)$ and $s_{1:n}(\cdot)$ are the hazard rate functions of $X_{1:n}$ and $Y_{1:n},$ respectively, we readily find
\begin{equation}\label{e20}
r_{1:n}\left(x\right)=\sum_{k=1}^n\left(\lambda_k+\alpha_k e^{\beta_k x}\right)
\end{equation} 
and
\begin{equation}\label{e21}
s_{1:n}\left(x\right)=\sum_{k=1}^n\left(\lambda_k^{*}+\alpha_k^{*} e^{\beta_k^{*} x}\right).
\end{equation} 

\hspace*{0.3 in} In the following results, stochastic comparison between minimum lifetimes are discussed with respect to hazard rate ordering, which strengthen Theorems~\ref{th1}~-~\ref{th3} for the independent case. 
\begin{t1}\label{th4}
Let $X_1,X_2,...,X_n$ be a set of independent random variables such that $X_i\sim $GM$\left(\alpha_i,\beta_i,\lambda_i\right)$, $i=1,2,...,n$. Let $Y_1,Y_2,...,Y_n$ be another set of independent random variables such that $Y_i\sim $GM$\left(\alpha_i^{*},\beta_i,\lambda_i\right)$, $i=1,2,...,n$. Suppose $\mbox{\boldmath $\alpha$}, \mbox{\boldmath $\alpha^*$}\in \mathcal{D}_+$ (or $\mathcal{E}_+$), $\mbox{\boldmath $\beta$}\in \mathcal{D}_+$. Then, 
$\mbox{\boldmath $\alpha$}\stackrel{m}{\succeq} \mbox{\boldmath $\alpha^*$}$ implies $X_{1:n}\leq_{hr}~(\geq_{hr})Y_{1:n}.$
\end{t1}
{\bf Proof:} Here, we need to prove that $r_{1:n}\left(x\right)=\sum_{k=1}^n\left(\lambda_k+\alpha_k e^{\beta_k x}\right)=\Psi_1(\mbox{\boldmath $\alpha$})$ is s-convex (s-concave) in $\mbox{\boldmath $\alpha$}$. Now, for $i\leq j$, if $\mbox{\boldmath $\alpha$}\in \mathcal{D}_+$ (or $\mathcal{E}_+$), $\mbox{\boldmath $\beta$}\in \mathcal{D}_+,$ then
\begin{equation}\label{e22}
\frac{\partial \Psi_1}{\partial \alpha_i}-\frac{\partial \Psi_1}{\partial \alpha_j}=x\left(e^{\beta_i x}-e^{\beta_j x}\right)\geq 0.
\end{equation}
Thus, by Lemma \ref{l6} (Lemma \ref{l7}), $\Psi_1(\mbox{\boldmath $\alpha$})$ is s-convex (s-concave) in $\mbox{\boldmath $\alpha$}$. This proves the result.  \hfill$\Box$
As the proofs of the next two results are quite similar to that of Theorem \ref{th4}, we do not present them here for the sake of brevity.
\begin{t1}\label{th5}
Let $X_1,X_2,...,X_n$ be a set of independent random variables such that $X_i\sim $GM$\left(\alpha_i,\beta_i,\lambda_i\right)$, $i=1,2,...,n$. Let $Y_1,Y_2,...,Y_n$ be another set of independent random variables such that $Y_i\sim $GM$\left(\alpha_i,\beta_i^{*},\lambda_i\right)$, $i=1,2,...,n$. Suppose $\mbox{\boldmath $\alpha$}$, $\mbox{\boldmath $\beta$}$ and $\mbox{\boldmath $\beta^*$}\in \mathcal{D}_+$ (or $\mathcal{E}_+$). Then, $\mbox{\boldmath $\beta$}\stackrel{m}{\succeq} \mbox{\boldmath $\beta^*$}$ implies $X_{1:n}\leq_{hr}~Y_{1:n}$.
\end{t1}

\begin{t1}\label{th6}
Let $X_1,X_2,...,X_n$ be a set of independent random variables such that $X_i\sim $GM$\left(\alpha_i,\beta_i,\lambda_i\right)$, $i=1,2,...,n$. Let $Y_1,Y_2,...,Y_n$ be another set of independent random variables such that $Y_i\sim $GM$\left(\alpha_i,\beta_i,\lambda_i^{*}\right)$, $i=1,2,...,n$. If $\sum_{k=1}^n \lambda_k\geq\sum_{k=1}^n \lambda_k^*,$ then $X_{1:n}\leq_{hr}Y_{1:n}$.
\end{t1}
It is useful to mention that for independent GM distributed lifetimes, Theorem \ref{th4} guarantees that the smallest lifetime of the first sample of insurees is smaller in terms of hazard rate order than that of the second sample of insurees with common heterogeneous exponential coefficient of mortality parameters and age-independent parameters, when the initial mortality of the first sample majorizes that of the second sample. Theorems \ref{th5} and \ref{th6} can be interpreted in a similar manner as well!\\
\hspace*{0.3 in} The natural question that arises is whether the results of Theorem \ref{th4} and Theorem \ref{th5} can be improved to $lr$ ordering. The counter-example given below shows that even for $n=2,$ no such ordering exists between $X_{1:n}$ and $Y_{1:n}$.
\begin{ce}
Let $\left(\alpha_1,\alpha_2\right)=(0.1,20.0)$, $\left(\alpha_1^*,\alpha_2^*\right)=(2.1,18.0)$, $\left(\beta_1,\beta_2\right)=(0.2,0.1)$ and $\left(\lambda_1,\lambda_2\right)=(0.6,0.5)$. Clearly, $\mbox{\boldmath $\alpha$}, \mbox{\boldmath $\alpha^*$}\in \mathcal{E}_+$, $\mbox{\boldmath $\beta$}\in \mathcal{D}_+$ and $\mbox{\boldmath $\alpha$}\stackrel{m}{\succeq} \mbox{\boldmath $\alpha^*$}$. But, Figure \ref{fig1}a) shows that there exists no $lr$ ordering between $X_{1:2}$ and $Y_{1:2}$.\\
Again, let $\left(\alpha_1,\alpha_2\right)=(20.0, 0.1)$, $\left(\beta_1,\beta_2\right)=(0.8,0.2)$, $\left(\beta_1^*,\beta_2^*\right)=(0.7,0.3)$ and $\left(\lambda_1,\lambda_2\right)=(0.5,0.6)$. Clearly, $\mbox{\boldmath $\alpha$}, \mbox{\boldmath $\beta$}$ and $\mbox{\boldmath $\beta^*$}\in \mathcal{D}_+$ and $\mbox{\boldmath $\beta$}\stackrel{m}{\succeq} \mbox{\boldmath $\beta^*$}$. But, Figure \ref{fig1}b) shows that there exists no $lr$ ordering between $X_{1:2}$ and $Y_{1:2}$.  
\begin{figure}[ht]
\centering
\begin{minipage}[b]{0.45\linewidth}
\includegraphics[height=7 cm]{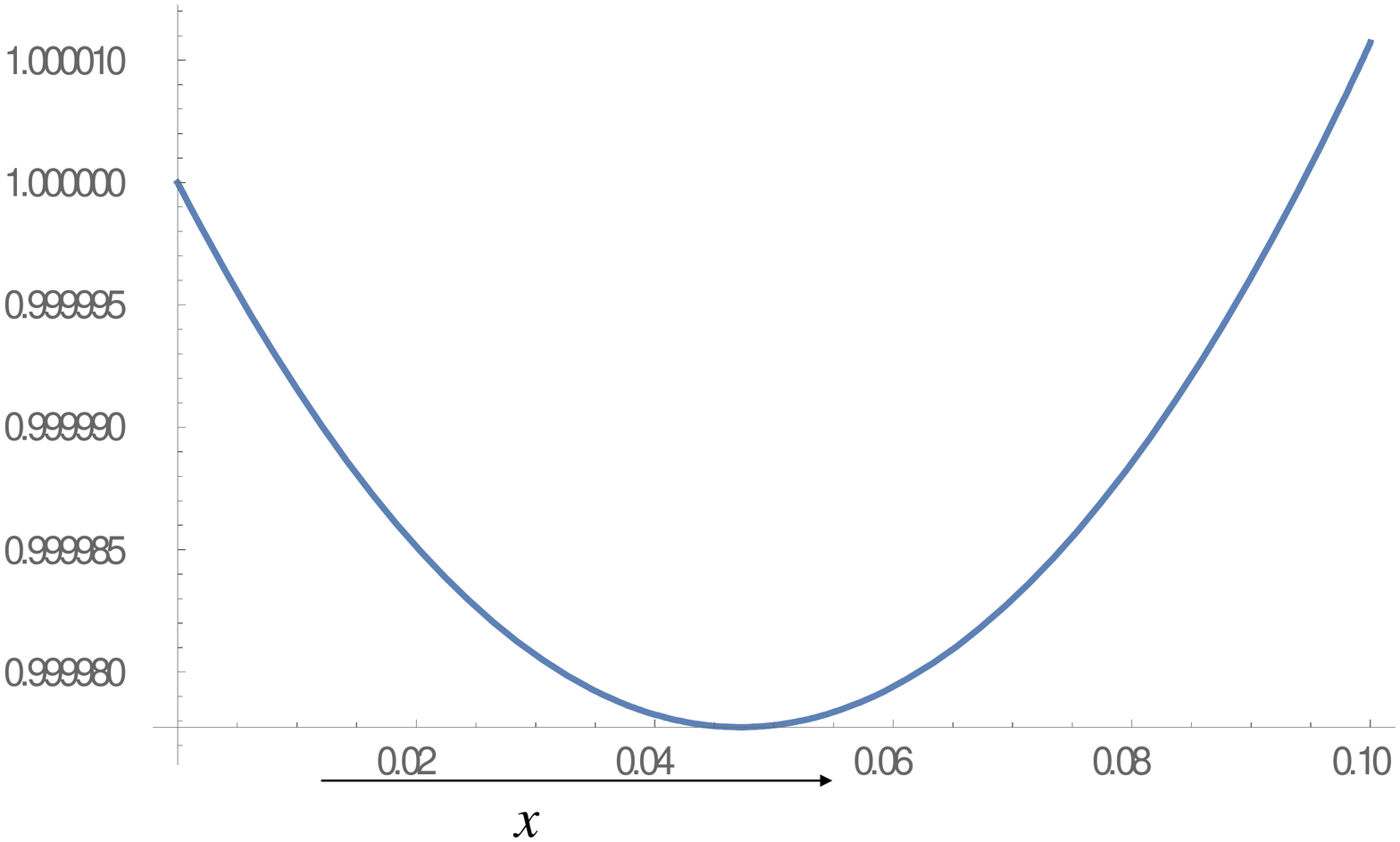}
\centering{$\left(i\right)$ Graph for $\mbox{\boldmath $\alpha$},\mbox{\boldmath $\alpha^*$}\in \mathcal{E}_+, \mbox{\boldmath $\beta$}\in \mathcal{D}_+$ }
\end{minipage}
\quad
\begin{minipage}[b]{0.45\linewidth}
\includegraphics[height=7 cm]{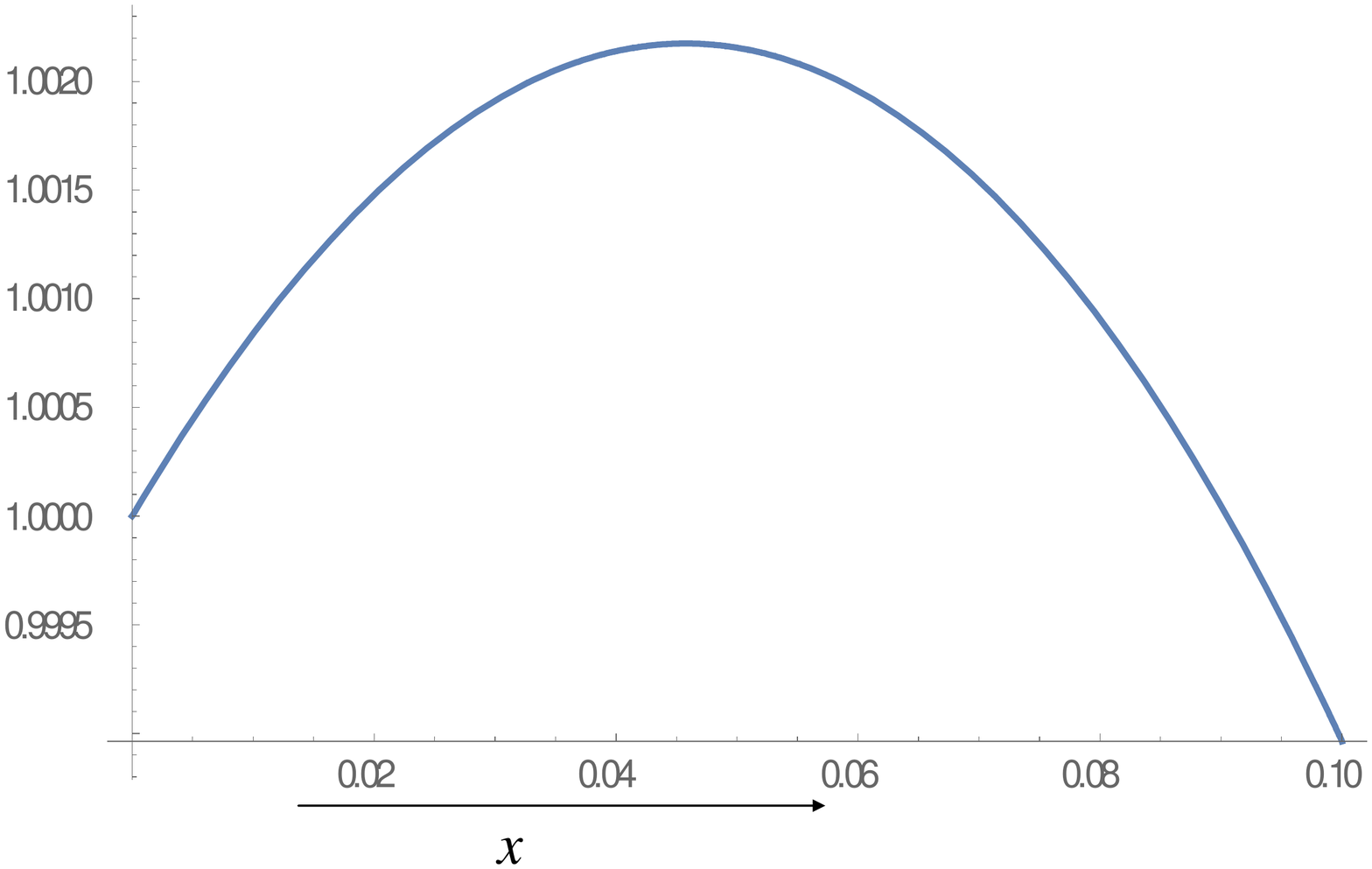}
\centering{$\left(ii\right)$ Graph for $\mbox{\boldmath $\alpha$}, \mbox{\boldmath $\beta$}, \mbox{\boldmath $\beta^*$}\in \mathcal{D}_+$}
\end{minipage}
\caption{\label{fig1}Graph of $\frac{g_{1:2}(x)}{h_{1:2}(x)}$} 
\end{figure}
\end{ce}

\hspace*{0.3 in} We now focus on the special case when the undderlying variables follow a multiple-outlier models. It is to be noted that a multiple-outlier model is a set of independent random variables $X_1,..., X_n$ of which $X_i\stackrel{st}{=}X, i=1,...,n_1,$ and $X_i\stackrel{st}{=}Y, i=n_1+1,...,n,$ where $1\leq n_1<n,$ and $X_i\stackrel{st}{=}X$ means that cdf of $X_i$ is same as that of $X.$ In other words, the set of independent random variables $X_1,..., X_n$ is said to constitute a multiple-outlier model if two sets of random variables $\left(X_1, X_2,\ldots, X_{n_1}\right)$ and $\left(X_{n_1+1}, X_{n_1+2},\ldots, X_{n_1+n_2}\right)$ (where $n_1+n_2=n$) are homogenous among themselves and heterogenous between themselves. For more details on multiple-outlier models, interested readers may refer to Balakrishnan~\cite{bala}, Kochar and Xu~(\cite{ko1}), Zhao and Balakrishnan~(\cite{zb11}), Kundu \emph{et al.}~\cite{kun1}, and the references therein. Then, the following results compare the minimum lifetimes of two groups of insurees from multiple-outlier model.

\begin{t1}\label{th9}
Let $\left\{X_1, X_2, \ldots, X_n\right\}$ and $\left\{Y_1, Y_2, \ldots, Y_n\right\}$ be two sets of independent random variables each following the multiple-outlier $GM$ model such that $X_i\sim $GM$\left(\alpha_1,\beta_1,\lambda_1\right)$ and $Y_i\sim $GM$\left(\alpha_1^{*},\beta_1,\lambda_1\right),$ for $i=1,2,\ldots,n_1,$ and $X_i\sim $GM$\left(\alpha_2,\beta_2,\lambda_2\right)$ and $Y_i\sim $GM$\left(\alpha_2^{*},\beta_2,\lambda_2\right),$ for $i=n_1+1,n_1+2,\ldots,n_1+n_2=n$. Then, $\alpha_1\leq (\geq) \alpha_2$, $\beta_1\geq \beta_2$ 
$$(\underbrace{\alpha_1,\alpha_1,\ldots,\alpha_1}_{n_1},\underbrace{\alpha_2,\alpha_2,\ldots,\alpha_2}_{n_2})\stackrel{m}{\succeq}
(\underbrace{\alpha_1^{*},\alpha_1^{*},\ldots,\alpha_1^{*}}_{n_1},\underbrace{\alpha_2^{*},\alpha_2^{*},\ldots,\alpha_2^{*}}_{n_2})\Rightarrow X_{1:n}\leq_{R-hr}~(\geq_{R-hr})Y_{1:n}.$$ 
\end{t1}
{\bf Proof:} To prove the result we have only to prove that $g(x)=\frac{s_{1:n}(x)}{r_{1:n}(x)}$ is decreasing (increasing) in $x$. Now, using the results in (\ref{e20}) and (\ref{e21}), we get  
$$g^{'}(x)\stackrel{sign}{=}\frac{\sum_{k=1}^n\left(\alpha_k\beta_k e^{\beta_k x}\right)}{\sum_{k=1}^n\left(\lambda_k+\alpha_k e^{\beta_k x}\right)}-\frac{\sum_{k=1}^n\left(\alpha_k^{*}\beta_k e^{\beta_k x}\right)}{\sum_{k=1}^n\left(\lambda_k+\alpha_k^* e^{\beta_k x}\right)}.$$
So, to prove the required result, we need to show that
$$\frac{\sum_{k=1}^n\left(\alpha_k\beta_k e^{\beta_k x}\right)}{\sum_{k=1}^n\left(\lambda_k+\alpha_k e^{\beta_k x}\right)}=\Psi_2(\mbox{\boldmath $\alpha$})~(say),$$ 
with $\mbox{\boldmath $\alpha$} =\left(\underbrace{\alpha_1,\ldots,\alpha_1}_{n_1},\underbrace{\alpha_2,\ldots,\alpha_2}_{n_2}\right),$ is s-concave (s-convex) in $\mbox{\boldmath $\alpha$}.$ \
Let $i\leq j$. Now, three cases need to be considered:\\
$Case (i):$ If $1\leq i,j\leq n_1$, $i.e.$, if $\alpha_i=\alpha_j=\alpha_1, \beta_i=\beta_j=\beta_1,$ and $\lambda_i=\lambda_j=\lambda_1$, then 
$$\frac{\partial \Psi_2}{\partial \alpha_i}-\frac{\partial \Psi_2}{\partial \alpha_j}=\frac{\partial \Psi_2}{\partial \alpha_1}-\frac{\partial \Psi_2}{\partial \alpha_1}=0;$$
$Case (ii):$ If $n_1+1\leq i,j\leq n$, $i.e.$, if $\alpha_i=\alpha_j=\alpha_2, \beta_i=\beta_j=\beta_2,$ and $\lambda_i=\lambda_j=\lambda_2$, then 
$$\frac{\partial \Psi_2}{\partial \alpha_i}-\frac{\partial \Psi_2}{\partial \alpha_j}=\frac{\partial \Psi_2}{\partial \alpha_2}-\frac{\partial \Psi_2}{\partial \alpha_2}=0;$$
$Case (iii):$ If $1\leq i\leq n_1$ and $n_1+1\leq j\leq n$, i.e. if $\alpha_i=\alpha_1$, $\alpha_j=\alpha_2$, $\beta_i=\beta_1$, $\beta_j=\beta_2$, and $\lambda_i=\lambda_1$ and $\lambda_j=\lambda_2$, then 
\begin{equation*}
\frac{\partial \Psi_2}{\partial \alpha_i}=\left(n_1\lambda_1+n_2\lambda_2\right)\beta_1 e^{\beta_1 x}+n_2\alpha_2(\beta_1-\beta_2)e^{(\beta_1+\beta_2) x}
\end{equation*}
and
\begin{equation*}
\frac{\partial \Psi_2}{\partial \alpha_j}=\left(n_1\lambda_1+n_2\lambda_2\right)\beta_2 e^{\beta_2 x}+n_1\alpha_1(\beta_2-\beta_1)e^{(\beta_1+\beta_2) x},
\end{equation*}
yielding 
\begin{equation}\label{1}
\frac{\partial \Psi_2}{\partial \alpha_i}-\frac{\partial \Psi_2}{\partial \alpha_j}\stackrel{sign}{=}\left(n_1\lambda_1+n_2\lambda_2\right)\left(\beta_1 e^{\beta_1 x}-\beta_2 e^{\beta_2 x}\right)+\left(n_1\alpha_1+n_2\alpha_2\right)(\beta_1-\beta_2)e^{(\beta_1+\beta_2) x}\geq 0,
\end{equation}
for $\beta_1\geq \beta_2.$ So, for $\alpha_1\leq (\geq)\alpha_2$, and by Lemma \ref{l7} (Lemma \ref{l6}), it can be concluded that $\Psi_2(\mbox{\boldmath $\alpha$})$ is s-concave (s-convex) in $\mbox{\boldmath $\alpha$}$. This proves the result.\hfill$\Box$\\
The following results can be proved in a similar way to Theorem \ref{th9} and so the proofs are not presented for the sake of brevity.
\begin{t1}\label{th10}
Let $\left\{X_1, X_2, \ldots, X_n\right\}$ and $\left\{Y_1, Y_2, \ldots, Y_n\right\}$ be two sets of independent random variables each following the multiple-outlier $GM$ model such that $X_i\sim $GM$\left(\alpha_1,\beta_1,\lambda_1\right)$ and $Y_i\sim $GM$\left(\alpha_1,\beta_1^{*},\lambda_1\right),$ for $i=1,2,\ldots,n_1,$ and $X_i\sim $GM$\left(\alpha_2,\beta_2,\lambda_2\right)$ and $Y_i\sim $GM$\left(\alpha_2,\beta_2^{*},\lambda_2\right),$ for $i=n_1+1,n_1+2,\ldots,n_1+n_2=n$. Then, $\alpha_1\geq \alpha_2,$ $\beta_1\geq \beta_2$ and
$$(\underbrace{\beta_1,\beta_1,\ldots,\beta_1}_{n_1},\underbrace{\beta_2,\beta_2,\ldots,\beta_2}_{n_2})\stackrel{m}{\succeq}
(\underbrace{\beta_1^{*},\beta_1^{*},\ldots,\beta_1^{*}}_{n_1},\underbrace{\beta_2^{*},\beta_2^{*},\ldots,\beta_2^{*}}_{n_2})\Rightarrow X_{1:n}\geq_{R-hr}Y_{1:n}.$$ 
\end{t1}
In the results above, it has been shown that for multiple-outlier model, $R-hr$ ordering exists between $X_{1:n}$ and $Y_{1:n}$ when $\mbox{\boldmath $\alpha$}$ majorizes  $\mbox{\boldmath $\alpha^*$}$ and  $\mbox{\boldmath $\beta$}$ majorizes  $\mbox{\boldmath $\beta^*$}$, keeping all the other parameter vectors equal. The next result shows that if $X_i\sim $GM$\left(\alpha_i,\beta_i,\lambda_i\right)$ and $Y_i\sim $GM$\left(\alpha_i,\beta_i,\lambda_i^{*}\right),$ for $i=1,2,\ldots,n$, then the same result holds under less restrictive conditions for a general model instead of multiple-outlier model.  
\begin{t1}\label{th11}
Let $\left\{X_1, X_2, \ldots, X_n\right\}$ and $\left\{Y_1, Y_2, \ldots, Y_n\right\}$ be two sets of independent random variables each following $GM$ model such that $X_i\sim $GM$\left(\alpha_i,\beta_i,\lambda_i\right)$ and $Y_i\sim $GM$\left(\alpha_i,\beta_i,\lambda_i^{*}\right),$ for $i=1,2,\ldots,n$. If  $\sum_{k=1}^n\lambda_k\geq\sum_{k=1}^n\lambda^*,$ then $X_{1:n}\geq_{R-hr}Y_{1:n}.$
\end{t1}
\begin{remark}
When $\mbox{\boldmath $\alpha $}$ majorizes $\mbox{\boldmath $\alpha^*$},$ depending on some conditions on $\mbox{\boldmath $\beta $}, \mbox{\boldmath $\alpha$}$, and $\mbox{\boldmath $\alpha^*$}$, Theorem \ref{th4} states that $X_{1:n}\geq_{hr}Y_{1:n}$, while Theorem \ref{th9} guarantees $X_{1:n}\leq_{R-hr}Y_{1:n}$. Similar results are obtained from Theorems \ref{th5} and \ref{th10} when $\mbox{\boldmath $\beta $}$ majorizes $\mbox{\boldmath $\beta^*$},$ depending on some conditions on $\mbox{\boldmath $\alpha $}, \mbox{\boldmath $\beta$}$, and $\mbox{\boldmath $\beta^*$}.$
\end{remark}

\section{Comparison of two sample maxima}
In this section, we compare maximum lifetimes of two groups of insurees when the GM distributed lifetimes have a dependent structure sharing Archimedean copula.\\
For $i=1,2,\ldots, n$, let $X_i\sim GM\left(\alpha_i, \beta_i, \lambda_i\right)$ and $Y_i\sim GM\left(\alpha_i^*, \beta_i^*, \lambda_i^*\right)$ be two sets of random variables sharing Archimedean copula having generators $\psi_1$ ($\phi_1=\psi_1^{-1}$) and $\psi_2$ ($\phi_2=\psi_2^{-1}$), respectively. Now, if $F_{n:n}(\cdot)$ and $G_{n:n}(\cdot)$ represent the distribution functions of $X_{n:n}$ and $Y_{n:n},$ respectively, then for all $x\geq 0$, we have 
\begin{equation*}
F_{n:n}\left(x\right)=\psi_1\left[\sum_{k=1}^n \phi_1\left(1-e^{-\lambda_k x-\frac{\alpha_k}{\beta_k}\left(e^{\beta_k x}-1\right)}\right)\right]
\end{equation*}
and
\begin{equation*}
G_{n:n}\left(x\right)=\psi_2\left[\sum_{k=1}^n \phi_2\left(1-e^{-\lambda_k^{*} x-\frac{\alpha_k^{*}}{\beta_k^{*}}\left(e^{\beta_k^{*} x}-1\right)}\right)\right].
\end{equation*}
In the following results, stochastic comparison between maximum order statistics with respect to usual stochastic ordering are discussed under different conditions. The next theorem discusses the ordering between $X_{n:n}$ and $Y_{n:n}$ when the parameter $\beta$ and parameter vector $\mbox{\boldmath $\lambda$}$ remain the same for the two distributions but $\mbox{\boldmath $\alpha^*$}$ is weakly supermajorized by $\mbox{\boldmath $\alpha$}$.
\begin{t1}\label{th12}
Let $X_1$, $X_2$, $\ldots X_n$ be a set of dependent random variables sharing Archimedean copula having generator $\psi_1,$ such that $X_i\sim GM\left(\alpha_i, \beta, \lambda_i\right)$, $i=1,2,\ldots n$. Further, let $Y_1$, $Y_2$, $\ldots Y_n$ be another set of dependent random variables sharing Archimedean copula having generator $\psi_2,$ such that $Y_i\sim GM\left(\alpha_i^*, \beta, \lambda_i\right)$, $i=1,2,\ldots n$. Assume that $\mbox{\boldmath $\alpha$},\mbox{\boldmath $\alpha^*$}, \mbox{\boldmath $\lambda$}\in \mathcal{D}_+ (\mathcal{E}_+)$. Suppose $\phi_2\circ\psi_1$ is super-additive and $\psi_1$ or $\psi_2$ is log-convex. Then, $\mbox{\boldmath $\alpha$}\stackrel{w}{\succeq}\mbox{\boldmath $\alpha^*$}$ implies $X_{n:n}\geq_{st}Y_{n:n}$.
\end{t1}
{\bf Proof:} For all $x\geq 0$, we have
\begin{equation*}
F_{n:n}\left(x\right)=\psi_1\left[\sum_{k=1}^n \phi_1\left(1-e^{-\lambda_k x-\alpha_k z}\right)\right]
\end{equation*}
and
\begin{equation*}
G_{n:n}\left(x\right)=\psi_2\left[\sum_{k=1}^n \phi_2\left(1-e^{-\lambda_k x-\alpha_k^*z}\right)\right],
\end{equation*}
where $z=\frac{e^{\beta x}-1}{\beta}$. Since $\phi_2\circ\psi_1$ is super-additive, then, by Lemma \ref{l11}, we have
\begin{equation}\label{eq4.1}
\psi_1\left[\sum_{k=1}^n\phi_1\left(1-e^{-\lambda_kx-\alpha_kz}\right)\right]\leq \psi_2\left[\sum_{k=1}^n\phi_2\left(1-e^{-\lambda_kx-\alpha_kz}\right)\right].
\end{equation}
Let $\psi_2\left[\sum_{k=1}^n\phi_2\left(1-e^{-\lambda_kx-\alpha_kz}\right)\right]=\Psi_3\left(\mbox{\boldmath $\alpha$}\right)$. Upon differentiating $\Psi_3\left(\mbox{\boldmath $\alpha$}\right)$ with respect to $\alpha_i$, we get
\begin{eqnarray}\frac{\partial\Psi_3}{\partial\alpha_i}&=&\psi_2^{'}\left[\sum_{k=1}^n \phi_2\left(1-e^{-\lambda_k x-\alpha_kz}\right)\right]\phi_2^{'}\left(1-e^{-\lambda_i x-\alpha_iz}\right)ze^{-\lambda_i x-\alpha_i z}\nonumber\\
&=&\psi_2^{'}\left[\sum_{k=1}^n \phi_2\left(1-e^{-\lambda_k x-\alpha_kz}\right)\right]\frac{\psi_2\left[\phi_2\left(1-e^{-\lambda_i x-\alpha_iz}\right)\right]}{\psi_2^{'}\left[\phi_2\left(1-e^{-\lambda_i x-\alpha_iz}\right)\right]}\frac{z}{e^{\lambda_i x+\alpha_iz}-1},\label{eq4.2}
\end{eqnarray}
which is greater than $0$, for all $x\geq 0$, yielding $\Psi_3\left(\mbox{\boldmath $\alpha$}\right)$ to be increasing in each $\alpha_i$. Since $\mbox{\boldmath $\alpha$}, \mbox{\boldmath $\lambda$}\in \mathcal{D}_+ (\mathcal{E}_+)$, for $i\leq j$, $\alpha_i\geq (\leq) \alpha_j$ and $\lambda_i\geq (\leq)\lambda_j$ gives 
\begin{equation}\label{eq4.3}
\frac{1}{e^{\lambda_i x+\alpha_iz}-1}\leq (\geq)\frac{1}{e^{\lambda_j x+\alpha_jz}-1}
\end{equation}
and
$$\phi_2\left(1-e^{-\lambda_i x-\alpha_iz}\right)\leq (\geq)\phi_2\left(1-e^{-\lambda_j x-\alpha_jz}\right).$$
Now, as $\psi_2(x)$ is log-convex in $x$ so that $\frac{\psi_2(x)}{\psi_2^{'}(x)}$ is decreasing in $x$, from the last inequality, we have,  for all $x\geq 0$,
\begin{equation}\label{eq4.4}
\frac{\psi_2\left(\phi_2\left(1-e^{-\lambda_i x-\alpha_iz}\right)\right)}{\psi_2^{'}\left(\phi_2\left(1-e^{-\lambda_i x-\alpha_i z}\right)\right)}\geq (\leq) \frac{\psi_2\left(\phi_2\left(1-e^{-\lambda_j x-\alpha_j z}\right)\right)}{\psi_2^{'}\left(\phi_2\left(1-e^{-\lambda_j x-\alpha_j z}\right)\right)}.
\end{equation}
Thus, using (\ref{eq4.3}) and (\ref{eq4.4}), we obtain 
\begin{eqnarray*}
\psi_2\left[\sum_{k=1}^n\phi_2\left(1-e^{-\lambda_kx-\alpha_kz}\right)\right]\frac{\psi_2\left(\phi_2\left(1-e^{-\lambda_i x-\alpha_iz}\right)\right)}{\psi_2^{'}\left(\phi_2\left(1-e^{-\lambda_i x-\alpha_i z}\right)\right)}\frac{z}{e^{\lambda_i x+\alpha_iz}-1}&\leq (\geq)&\\
\psi_2\left[\sum_{k=1}^n\phi_2\left(1-e^{-\lambda_kx-\alpha_kz}\right)\right]\frac{\psi_2\left(\phi_2\left(1-e^{-\lambda_j x-\alpha_jz}\right)\right)}{\psi_2^{'}\left(\phi_2\left(1-e^{-\lambda_j x-\alpha_j z}\right)\right)}\frac{z}{e^{\lambda_j x+\alpha_jz}-1},&&
\end{eqnarray*}
which, by (\ref{eq4.2}), gives $\frac{\partial\Psi_3}{\partial\alpha_i}-\frac{\partial\Psi_3}{\partial\alpha_j}\leq (\geq)0$. Thus, by Lemmas \ref{l6} and \ref{l7}, we get $\Psi_3\left(\mbox{\boldmath $\alpha$}\right)$ to be s-concave in $\mbox{\boldmath $\alpha$}$. So, through the fact that $\Psi_3\left(\mbox{\boldmath $\alpha$}\right)$ is increasing in each $\alpha_i$, using Lemma \ref{l33} and (\ref{eq4.1}), the result follows.\hfill$\Box$\\
The following results discuss the case when there exists weak super-majorization ordering between the vectors $\mbox{\boldmath $\lambda$}$ and $\mbox{\boldmath $\lambda^*$}$ under two situations: (i) when the parameter $\beta$ and the parameter vector $\mbox{\boldmath $\alpha$}$ are the same for the two distributions, and (ii) when the parameter $\alpha$ and the parameter vector $\mbox{\boldmath $\beta$}$ are the same for the two distributions. These results can be proved along he same line as Theorem \ref{th12} and so are not presented here. 
\begin{t1}\label{th13}
Let $X_1$, $X_2$, $\ldots X_n$ be a set of dependent random variables sharing Archimedean copula having generator $\psi_1,$ such that $X_i\sim GM\left(\alpha_i, \beta, \lambda_i\right)$, $i=1,2,\ldots n$. Also, let $Y_1$, $Y_2$, $\ldots Y_n$ be another set of dependent random variables sharing Archimedean copula having generator $\psi_2,$ such that $Y_i\sim GM\left(\alpha_i, \beta, \lambda_i^*\right)$, $i=1,2,\ldots n$. Assume that $\mbox{\boldmath $\alpha$},\mbox{\boldmath $\lambda$}, \mbox{\boldmath $\lambda^*$}\in \mathcal{D}_+ (\mathcal{E}_+)$. Further, suppose $\phi_2\circ\psi_1$ is super-additive and $\psi_1$ or $\psi_2$ is log-convex. Then, $\mbox{\boldmath $\lambda$}\stackrel{w}{\succeq}\mbox{\boldmath $\lambda^*$}$ implies $X_{n:n}\geq_{st}Y_{n:n}$.
\end{t1}
\begin{t1}\label{th14}
Let $X_1$, $X_2$, $\ldots X_n$ be a set of dependent random variables sharing Archimedean copula having generator $\psi_1,$ such that $X_i\sim GM\left(\alpha, \beta_i, \lambda_i\right)$, $i=1,2,\ldots n$. Also, let $Y_1$, $Y_2$, $\ldots Y_n$ be another set of dependent random variables sharing Archimedean copula having generator $\psi_2,$ such that $Y_i\sim GM\left(\alpha, \beta_i, \lambda_i^*\right)$, $i=1,2,\ldots n$. Assume that $\mbox{\boldmath $\beta$},\mbox{\boldmath $\lambda$}, \mbox{\boldmath $\lambda^*$}\in \mathcal{D}_+ (\mathcal{E}_+)$. Further, suppose $\phi_2\circ\psi_1$ is super-additive and $\psi_1$ or $\psi_2$ is log-convex. Then, $\mbox{\boldmath $\lambda$}\stackrel{w}{\succeq}\mbox{\boldmath $\lambda^*$}$ implies $X_{n:n}\geq_{st}Y_{n:n}$.
\end{t1}
In the following theorem, ordering between $X_{n:n}$ and $Y_{n:n}$ is presented when the parameter $\alpha$ and parameter vector $\mbox{\boldmath $\lambda$}$ are the same for the two distributions, but there exists majorization ordering between the parameter vectors $\mbox{\boldmath $\beta$}$ and $\mbox{\boldmath $\beta^*$}$.
\begin{t1}\label{th15}
Let $X_1$, $X_2$, $\ldots X_n$ be a set of dependent random variables sharing Archimedean copula having generator $\psi_1,$ such that $X_i\sim GM\left(\alpha, \beta_i, \lambda_i\right)$, $i=1,2,\ldots n$. Also, let $Y_1$, $Y_2$, $\ldots Y_n$ be another set of dependent random variables sharing Archimedean copula having generator $\psi_2,$ such that $Y_i\sim GM\left(\alpha, \beta_i^*, \lambda_i\right)$, $i=1,2,\ldots n$. Assume that $\mbox{\boldmath $\beta$},\mbox{\boldmath $\beta^*$}, \mbox{\boldmath $\lambda$}\in \mathcal{D}_+ (\mathcal{E}_+)$. Further, suppose $\phi_2\circ\psi_1$ is super-additive and $\psi_1$ or $\psi_2$ is log-convex. Then, $\frac{1}{\mbox{\boldmath $\beta$}}\succeq_w\frac{1}{\mbox{\boldmath $\beta^*$}}$ implies $X_{n:n}\geq_{st}Y_{n:n}$, where $\frac{1}{\mbox{\boldmath $\beta$}}\equiv\left(\frac{1}{\beta_1}, \frac{1}{\beta_2},\ldots\frac{1}{\beta_n}\right)$.
\end{t1}
{\bf Proof:} For all $x\geq 0$, we have
\begin{equation*}
F_{n:n}\left(x\right)=\psi_1\left[\sum_{k=1}^n \phi_1\left(1-e^{-\lambda_k x-\alpha z_k}\right)\right]
\end{equation*}
and
\begin{equation*}
G_{n:n}\left(x\right)=\psi_2\left[\sum_{k=1}^n \phi_2\left(1-e^{-\lambda_k x-\alpha z_k^*}\right)\right],
\end{equation*}
where $z_k=\frac{e^{\beta_k x}-1}{\beta_k}$ and $z_k^*=\frac{e^{\beta_k^* x}-1}{\beta_k^*}$. Since $\phi_2\circ\psi_1$ is super-additive, then, by Lemma \ref{l11}, we have
\begin{equation}\label{eq4.5}
\psi_1\left[\sum_{k=1}^n\phi_1\left(1-e^{-\lambda_kx-\alpha z_k}\right)\right]\leq \psi_2\left[\sum_{k=1}^n\phi_2\left(1-e^{-\lambda_kx-\alpha z_k}\right)\right].
\end{equation}
Let $\psi_2\left[\sum_{k=1}^n\phi_2\left(1-e^{-\lambda_k x-\alpha z_k}\right)\right]=\Psi_4\left(\mbox{\boldmath $z$}\right)$. Upon differentiating $\Psi_4\left(\mbox{\boldmath $z$}\right)$ with respect to $z_i$, we get
\begin{eqnarray}\frac{\partial\Psi_4}{\partial z_i}&=&\psi_2^{'}\left[\sum_{k=1}^n \phi_2\left(1-e^{-\lambda_k x-\alpha z_k}\right)\right]\phi_2^{'}\left(1-e^{-\lambda_i x-\alpha z_i}\right)\alpha e^{-\lambda_i x-\alpha z_i}\nonumber\\
&=&\psi_2^{'}\left[\sum_{k=1}^n \phi_2\left(1-e^{-\lambda_k x-\alpha z_k}\right)\right]\frac{\psi_2\left[\phi_2\left(1-e^{-\lambda_i x-\alpha z_i}\right)\right]}{\psi_2^{'}\left[\phi_2\left(1-e^{-\lambda_i x-\alpha z_i}\right)\right]}\frac{\alpha}{e^{\lambda_i x+\alpha z_i}-1},\label{eq4.6}
\end{eqnarray}
which is greater than $0$, for all $x\geq 0$, implying that $\Psi_4\left(\mbox{\boldmath $z$}\right)$ is increasing in each $z_i$. Because each $z_i$ is increasing function of $\beta_i$, $\mbox{\boldmath $\beta$}\in \mathcal{D}_+ (\mathcal{E}_+)$ implies $\mbox{\boldmath $z$}\in \mathcal{D}_+ (\mathcal{E}_+)$. So, as $\mbox{\boldmath $z$}, \mbox{\boldmath $\lambda$}\in \mathcal{D}_+ (\mathcal{E}_+)$, for $i\leq j$, $z_i\geq (\leq) z_j$ and $\lambda_i\geq (\leq)\lambda_j$ gives 
\begin{equation}\label{eq4.7}
\frac{1}{e^{\lambda_i x+\alpha z_i}-1}\leq (\geq)\frac{1}{e^{\lambda_j x+\alpha z_j}-1}
\end{equation}
and
$$\phi_2\left(1-e^{-\lambda_i x-\alpha z_i}\right)\leq (\geq)\phi_2\left(1-e^{-\lambda_j x-\alpha z_j}\right).$$
Now, as $\psi_2(x)$ is log-convex in $x$ so that $\frac{\psi_2(x)}{\psi_2^{'}(x)}$ is decreasing in $x$, from the last inequality, we have, for all $x\geq 0$,
\begin{equation}\label{eq4.8}
\frac{\psi_2\left(\phi_2\left(1-e^{-\lambda_i x-\alpha z_i}\right)\right)}{\psi_2^{'}\left(\phi_2\left(1-e^{-\lambda_i x-\alpha z_i}\right)\right)}\geq (\leq) \frac{\psi_2\left(\phi_2\left(1-e^{-\lambda_j x-\alpha z_j}\right)\right)}{\psi_2^{'}\left(\phi_2\left(1-e^{-\lambda_j x-\alpha z_j}\right)\right)}.
\end{equation}
Thus, using (\ref{eq4.7}) and (\ref{eq4.8}), we obtain 
\begin{eqnarray*}
\psi_2\left[\sum_{k=1}^n\phi_2\left(1-e^{-\lambda_kx-\alpha z_k}\right)\right]\frac{\psi_2\left(\phi_2\left(1-e^{-\lambda_i x-\alpha z_i}\right)\right)}{\psi_2^{'}\left(\phi_2\left(1-e^{-\lambda_i x-\alpha z_i}\right)\right)}\frac{\alpha}{e^{\lambda_i x+\alpha z_i}-1}&\leq (\geq)&\\
\psi_2\left[\sum_{k=1}^n\phi_2\left(1-e^{-\lambda_kx-\alpha z_k}\right)\right]\frac{\psi_2\left(\phi_2\left(1-e^{-\lambda_j x-\alpha z_j}\right)\right)}{\psi_2^{'}\left(\phi_2\left(1-e^{-\lambda_j x-\alpha z_j}\right)\right)}\frac{\alpha}{e^{\lambda_j x+\alpha z_j}-1},&&
\end{eqnarray*}
which, by (\ref{eq4.6}), gives $\frac{\partial\Psi_4}{\partial z_i}-\frac{\partial\Psi_4}{\partial z_j}\leq (\geq)0$. Thus, by Lemmas \ref{l6} and \ref{l7}, we have $\Psi_4\left(\mbox{\boldmath $z$}\right)$ to be s-concave in $\mbox{\boldmath $z$}$. So, using the fact that $\Psi_4\left(\mbox{\boldmath $z$}\right)$ is increasing in each $z$, using Lemma \ref{l33}, it can be shown that 
\begin{equation}\label{eq4.9}
\mbox{\boldmath $z$}\stackrel{w}{\succeq}\mbox{\boldmath $z^*$}\Longrightarrow\Psi_4\left(\mbox{\boldmath $z$}\right)\leq\Psi_4\left(\mbox{\boldmath $z^*$}\right).
\end{equation}
 Now, by taking $\beta_i=-\frac{1}{\nu_i}$ and differentiating $z_i$ with respect to $\nu_i,$ we get 
$\frac{\partial z_i}{\partial\nu_i}=1-e^{-x/\nu_i}-\frac{x}{\nu_i}e^{-x/\nu_i}$, the value of which is equal to $0$ at $x=0$. Again, for all $x\geq 0$, we have
$$\frac{\partial}{\partial x}\left(1-e^{-x/\nu_i}-\frac{x}{\nu_i}e^{-x/\nu_i}\right)=\frac{x}{\nu_i^2}e^{-x/\nu_i}\geq 0,$$
implying $1-e^{-x/\nu_i}-\frac{x}{\nu_i}e^{-x/\nu_i}$ is increasing in $x$. So, for all $x\geq 0$, since $\frac{\partial z_i}{\partial\nu_i}\geq 0$, it can be concluded that $z_i$ is increasing in each $\nu_i$. Also, since $\frac{\partial^2z_i}{\partial\nu_i^2}=-\frac{x^2}{\nu_i^3}e^{-x/\nu_i}\leq 0$, $z_i$ is concave in $\nu_i$. Thus, by Theorem A2(ii) of Marshall \emph{et al.}~\cite{Maol} (page-167), it can be written that 
\begin{equation}\label{eq4.10}
 \mbox{\boldmath $\nu$}\stackrel{w}{\succeq}\mbox{\boldmath $\nu^*$}\Longrightarrow\mbox{\boldmath $z$}\stackrel{w}{\succeq}\mbox{\boldmath $z^*$},
\end{equation}
where $\mbox{\boldmath $\nu$}\equiv\left(-\frac{1}{\beta_1},-\frac{1}{\beta_2}\ldots,-\frac{1}{\beta_n}\right)$. Now, upon noticing the fact that
$$\frac{1}{\mbox{\boldmath $\beta$}}\succeq_{w}\frac{1}{\mbox{\boldmath $\beta^*$}}\Longleftrightarrow -\frac{1}{\mbox{\boldmath $\beta$}}\stackrel{w}{\succeq} -\frac{1}{\mbox{\boldmath $\beta^*$}},$$ 
using (\ref{eq4.9}) and (\ref{eq4.10}), it can be observed that 
$$\frac{1}{\mbox{\boldmath $\beta$}}\succeq_{w}\frac{1}{\mbox{\boldmath $\beta^*$}}\Longrightarrow\mbox{\boldmath $\nu$}\stackrel{w}{\succeq}\mbox{\boldmath $\nu^*$}\Longrightarrow\mbox{\boldmath $z$}\stackrel{w}{\succeq}\mbox{\boldmath $z^*$}\Longrightarrow\Psi_4\left(\mbox{\boldmath $z$}\right)\leq\Psi_4\left(\mbox{\boldmath $z^*$}\right).$$
So, by (\ref{eq4.5}), the result is proved.\hfill$\Box$\\
The natural question that arises here is whether any stochastic ordering exists between $X_{n:n}$ and $Y_{n:n}$ when $\lambda$ is kept fixed but there exists majorization ordering between either $\mbox{\boldmath $\alpha$}$ and $\mbox{\boldmath $\alpha^*$}$ or $\mbox{\boldmath $\beta$}$ and $\mbox{\boldmath $\beta^*$}$, while all the other parameter vectors remain the same. The following counter-examples show that no such ordering exists, even if the week super-majorization ordering is replaced by majorization ordering.
\begin{ce}\label{ce1}
Let $\lambda=0.6$, $\mbox{\boldmath $\beta$}=\left(2,1\right)$, $\mbox{\boldmath $\alpha$}=\left(0.2, 0.1\right)$ and $\mbox{\boldmath $\alpha^*$}=\left(0.18, 0.12\right)$. Then, $\mbox{\boldmath $\beta$}, \mbox{\boldmath $\alpha$}, \mbox{\boldmath $\alpha^*$}\in\mathcal{D}_+$ and $\mbox{\boldmath $\alpha$}\stackrel{m}{\succeq}\mbox{\boldmath $\alpha^*$}$. But, Figure \ref{fig2}(a) shows that there exists no stochastic ordering between $X_{n:n}$ and $Y_{n:n}$. 
\end{ce}

\begin{ce}\label{ce2}
Let $\lambda=0.02$, $\mbox{\boldmath $\beta$}=\left(\frac{1}{2},1\right)$, $\mbox{\boldmath $\beta^*$}=\left(\frac{1}{1.6}, \frac{1}{1.4}\right)$ and $\mbox{\boldmath $\alpha$}=\left(0.1, 0.2\right)$. Then, $\mbox{\boldmath $\beta$}, \mbox{\boldmath $\beta^*$}, \mbox{\boldmath $\alpha$}\in\mathcal{E}_+$ and $\frac{1}{\mbox{\boldmath $\beta$}}\stackrel{m}{\succeq}\frac{1}{\mbox{\boldmath $\beta^*$}}$. But, Figure \ref{fig2}(b) shows that there exists no stochastic ordering between $X_{n:n}$ and $Y_{n:n}$. 
\begin{figure}[ht]
\centering
\begin{minipage}[b]{0.48\linewidth}
\includegraphics[height=7 cm]{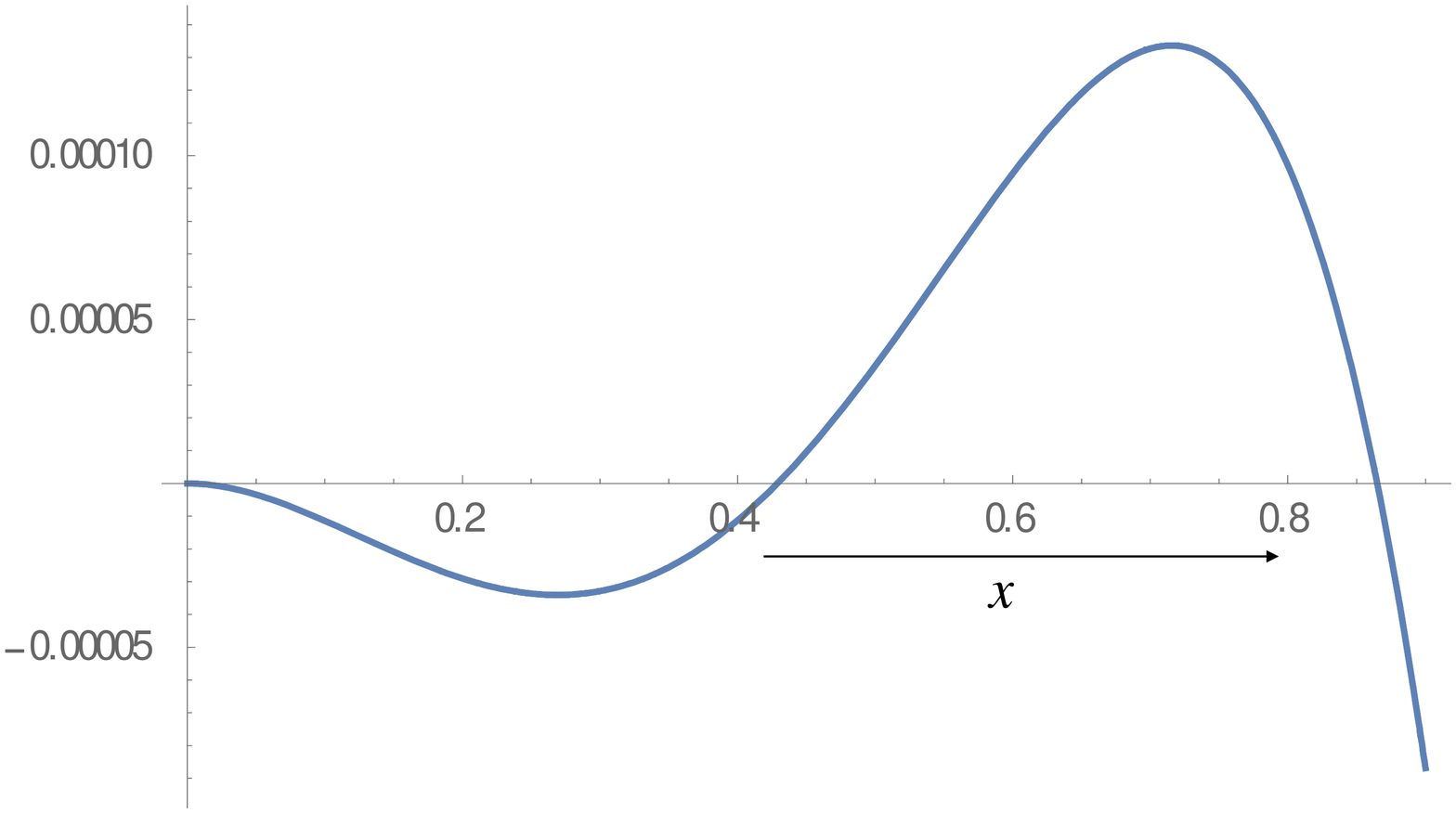}
\centering{(a) Counter-example \ref{ce1} }
\end{minipage}
\quad
\begin{minipage}[b]{0.48\linewidth}
\includegraphics[height=7 cm]{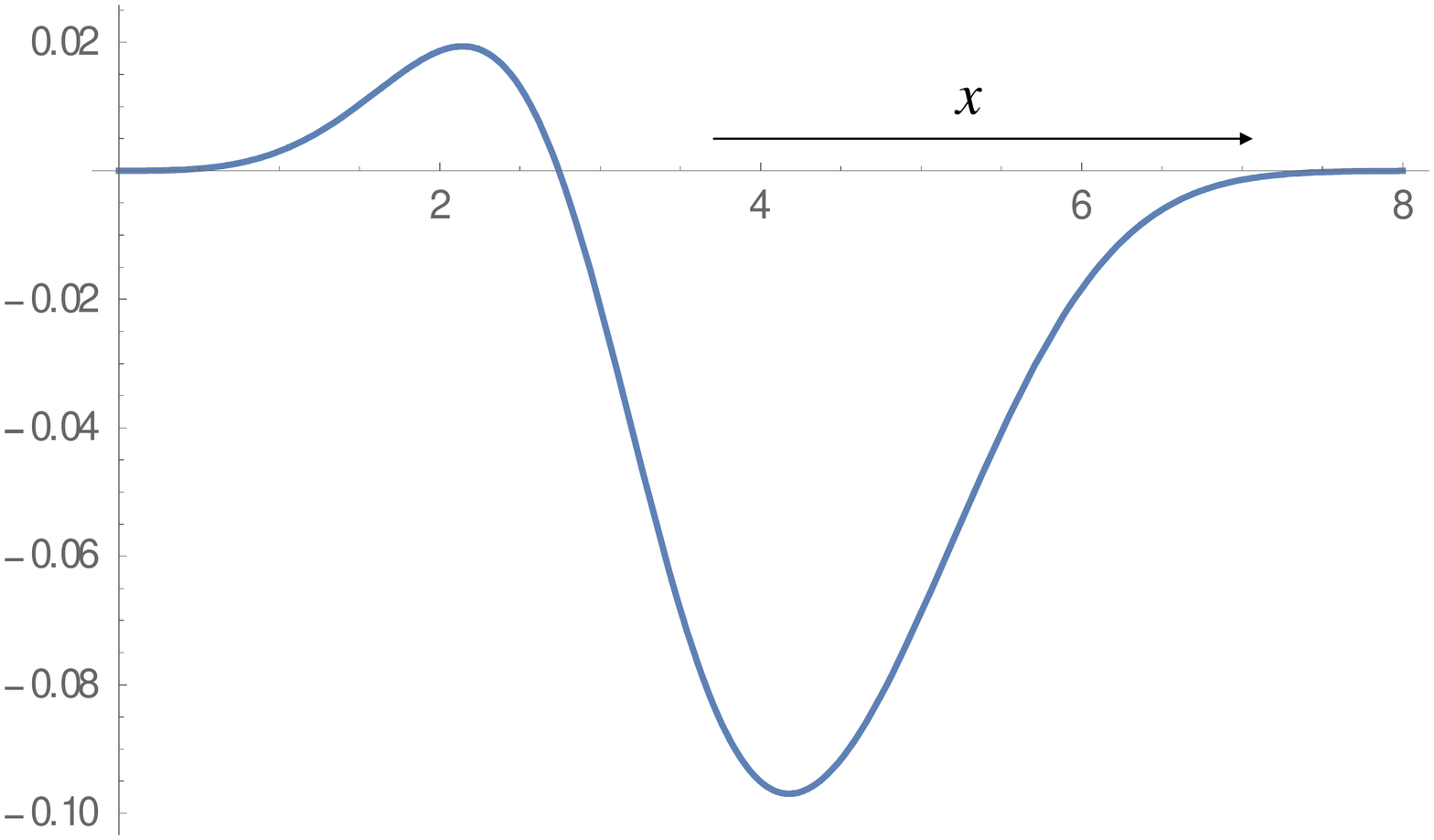}
\centering{(b) Counter-example \ref{ce2}}
\end{minipage}
\caption{\label{fig2}Graph of $F_{n:n}(x)-G_{n:n}(x)$} 
\end{figure}
\end{ce}

\begin{remark}\label{re1}
From Theorems \ref{th12}-\ref{th15} and Counter-examples \ref{ce1} and \ref{ce2}, we can conclude the following:
 \begin{enumerate}
\item[(i)] If $\beta$ is constant and 
\begin{enumerate}
\item[(a)] $\mbox{\boldmath $\lambda$}$ is the same for the two distributions, then $\mbox{\boldmath $\alpha$}\stackrel{w}{\succeq}\mbox{\boldmath $\alpha^*$}$ implies $X_{n:n}\geq_{st}Y_{n:n},$
\item[(b)] $\mbox{\boldmath $\alpha$}$ is the same for the two distributions, then $\mbox{\boldmath $\lambda$}\stackrel{w}{\succeq}\mbox{\boldmath $\lambda^*$}$ implies $X_{n:n}\geq_{st}Y_{n:n};$
\end{enumerate}
\item[(ii)] If $\alpha$ is constant and 
\begin{enumerate}
\item[(a)] $\mbox{\boldmath $\beta$}$ is the same for the two distributions, then $\mbox{\boldmath $\lambda$}\stackrel{w}{\succeq}\mbox{\boldmath $\lambda^*$}$ implies $X_{n:n}\geq_{st}Y_{n:n},$
\item[(b)] $\mbox{\boldmath $\lambda$}$ is the same for the two distributions, then $\frac{1}{\mbox{\boldmath $\beta$}}\succeq_{w}\frac{1}{\mbox{\boldmath $\beta^*$}}$ implies $X_{n:n}\geq_{st}Y_{n:n};$
\end{enumerate}
\item[(iii)] If $\lambda$ is constant and 
\begin{enumerate}
\item[(a)] $\mbox{\boldmath $\beta$}$ is the same for the two distributions, then $\mbox{\boldmath $\alpha$}\stackrel{w}{\succeq}\mbox{\boldmath $\alpha^*$}$ implies no st ordering between $X_{n:n}$ and $Y_{n:n},$
\item[(b)] $\mbox{\boldmath $\alpha$}$ is the same for the two distributions, then $\frac{1}{\mbox{\boldmath $\beta$}}\succeq_{w}\frac{1}{\mbox{\boldmath $\beta^*$}}$ implies no st ordering between $X_{n:n}$ and $Y_{n:n}$.
\end{enumerate}
\end{enumerate}
\end{remark}
The following counter-example shows that the results of Theorems \ref{th12}-\ref{th15} cannot be improved further to reversed hazard rate orderings.
\begin{ce}\label{ce3}
\begin{enumerate}
\item[(i)] For $\beta=2$, $\mbox{\boldmath $\lambda$}=\left(0.6,0.5\right)$, $\mbox{\boldmath $\alpha$}=\left(20, 0.1\right)$ and $\mbox{\boldmath $\alpha$}=\left(18, 2.1\right)$, although $\mbox{\boldmath $\alpha$}\stackrel{m}{\succeq}\mbox{\boldmath $\alpha^*$}$, Figure \ref{fig3}(a) shows that there exists no reversed hazard rate ordering between $X_{n:n}$ and $Y_{n:n};$
\item[(ii)] For $\beta=0.2$, $\mbox{\boldmath $\alpha$}=\left(0.02,0.01\right)$, $\mbox{\boldmath $\lambda$}=\left(0.07, 0.05\right)$ and $\mbox{\boldmath $\lambda^*$}=\left(0.06, 0.06\right)$, although $\mbox{\boldmath $\lambda$}\stackrel{m}{\succeq}\mbox{\boldmath $\lambda^*$}$, Figure \ref{fig3}(b) shows that there exists no reversed hazard rate ordering between $X_{n:n}$ and $Y_{n:n};$
\item[(iii)] For $\alpha=0.02$, $\mbox{\boldmath $\beta$}=\left(0.2,0.1\right)$, $\mbox{\boldmath $\lambda$}=\left(0.07, 0.05\right)$ and $\mbox{\boldmath $\lambda^*$}=\left(0.06, 0.06\right)$, although $\mbox{\boldmath $\lambda$}\stackrel{m}{\succeq}\mbox{\boldmath $\lambda^*$}$, Figure \ref{fig3}(c) shows that there exists no reversed hazard rate ordering between $X_{n:n}$ and $Y_{n:n};$
\item[(iv)] For $\alpha=0.02$, $\mbox{\boldmath $\lambda$}=\left(0.05,0.07\right)$, $\frac{1}{\mbox{\boldmath $\beta$}}=\left(0.3, 0.1\right)$ and $\frac{1}{\mbox{\boldmath $\beta^*$}}=\left(0.2, 0.2\right)$, although $\frac{1}{\mbox{\boldmath $\beta$}}\stackrel{m}{\succeq}\frac{1}{\mbox{\boldmath $\beta^*$}}$, Figure \ref{fig3}(d) shows that there exists no reversed hazard rate ordering between $X_{n:n}$ and $Y_{n:n}$.
\end{enumerate}
\begin{figure}[ht]
\centering
\begin{minipage}[b]{0.48\linewidth}
\includegraphics[height=7 cm]{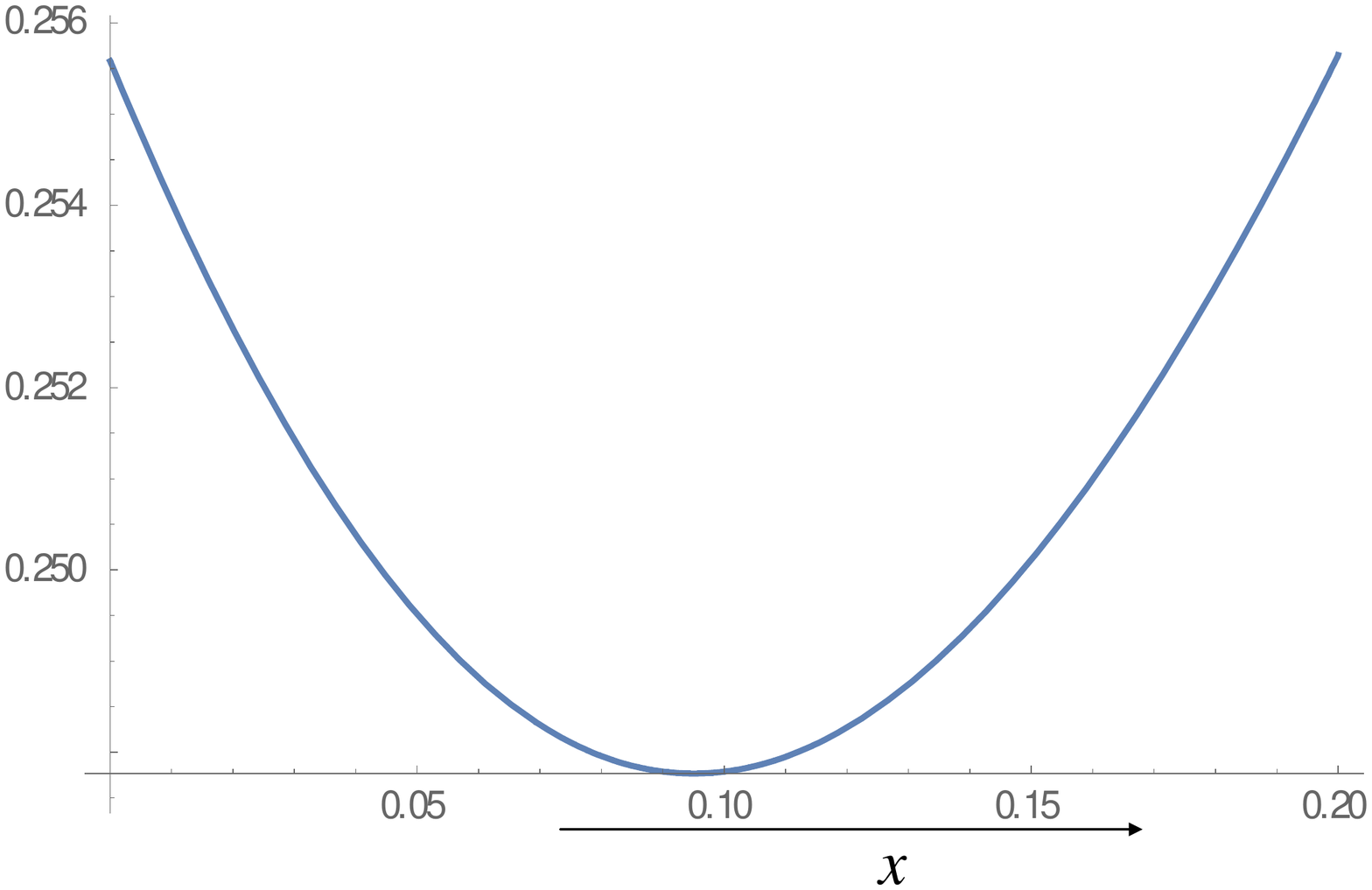}
\centering{(a) Counter-example \ref{ce3}(i) }
\end{minipage}
\quad
\begin{minipage}[b]{0.48\linewidth}
\includegraphics[height=7 cm]{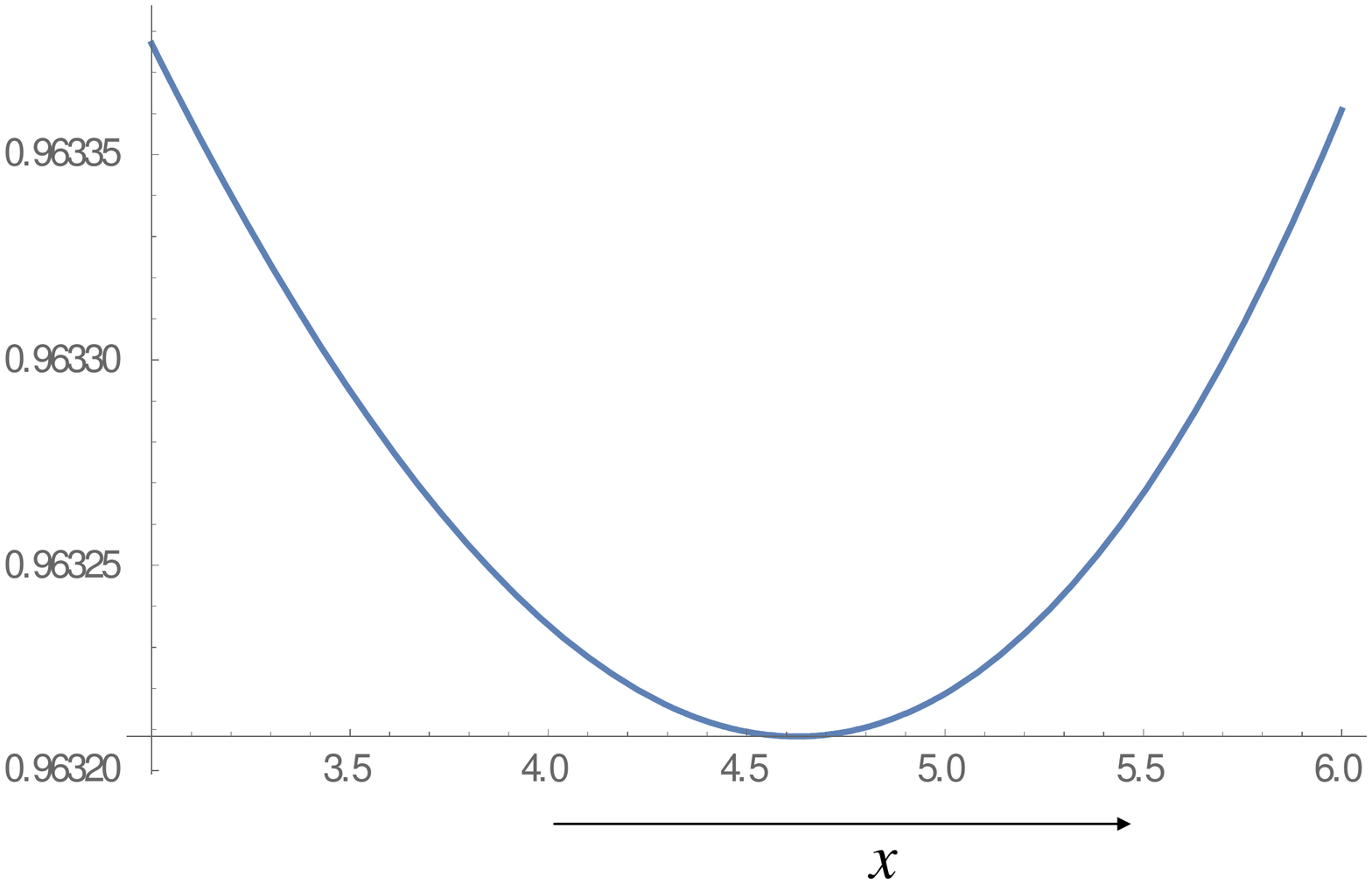}
\centering{(b) Counter-example \ref{ce3}(ii)}
\end{minipage}
\quad
\begin{minipage}[b]{0.48\linewidth}
\includegraphics[height=7 cm]{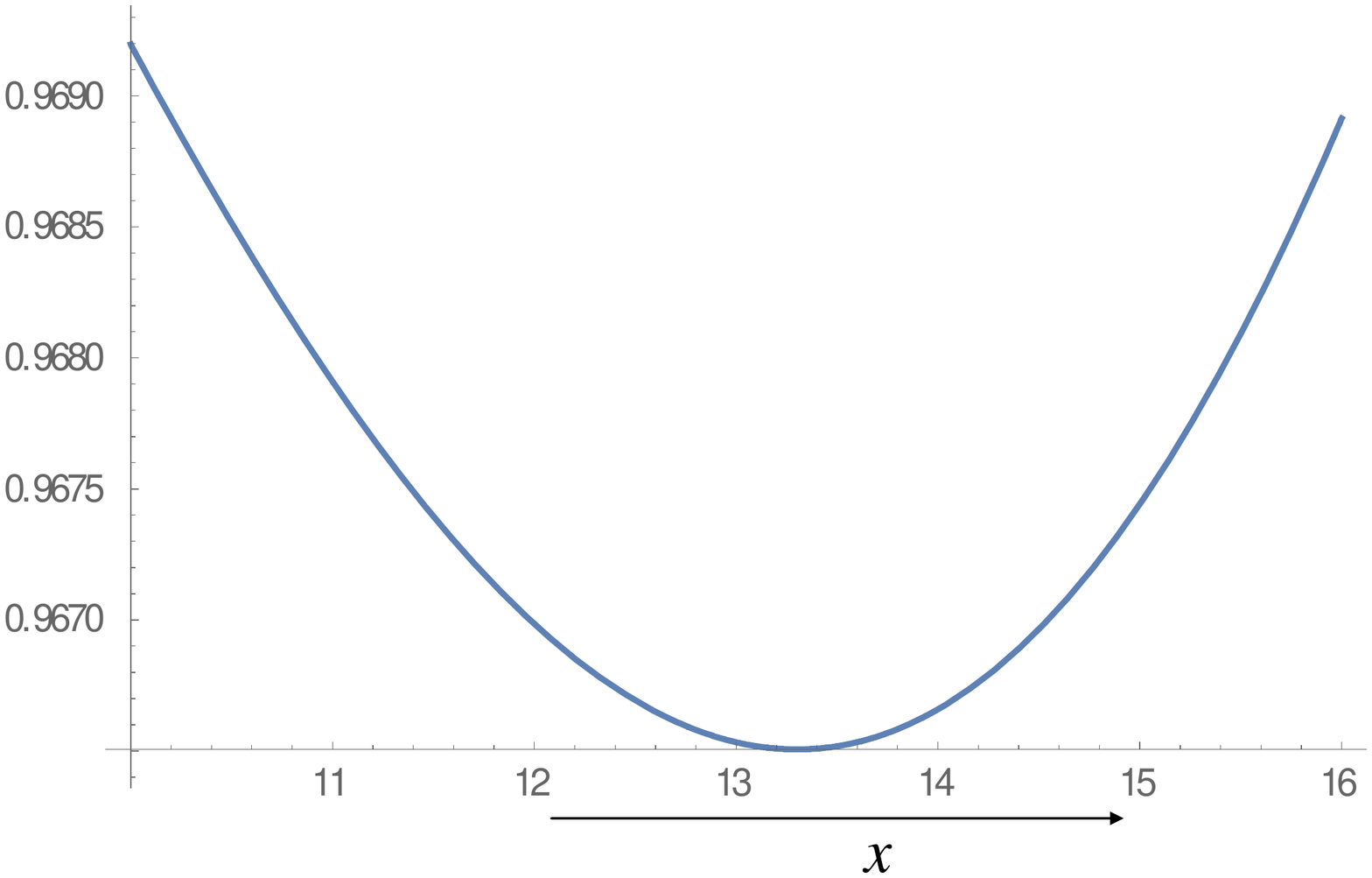}
\centering{(c) Counter-example \ref{ce3}(iii)}
\end{minipage}
\quad
\begin{minipage}[b]{0.48\linewidth}
\includegraphics[height=7 cm]{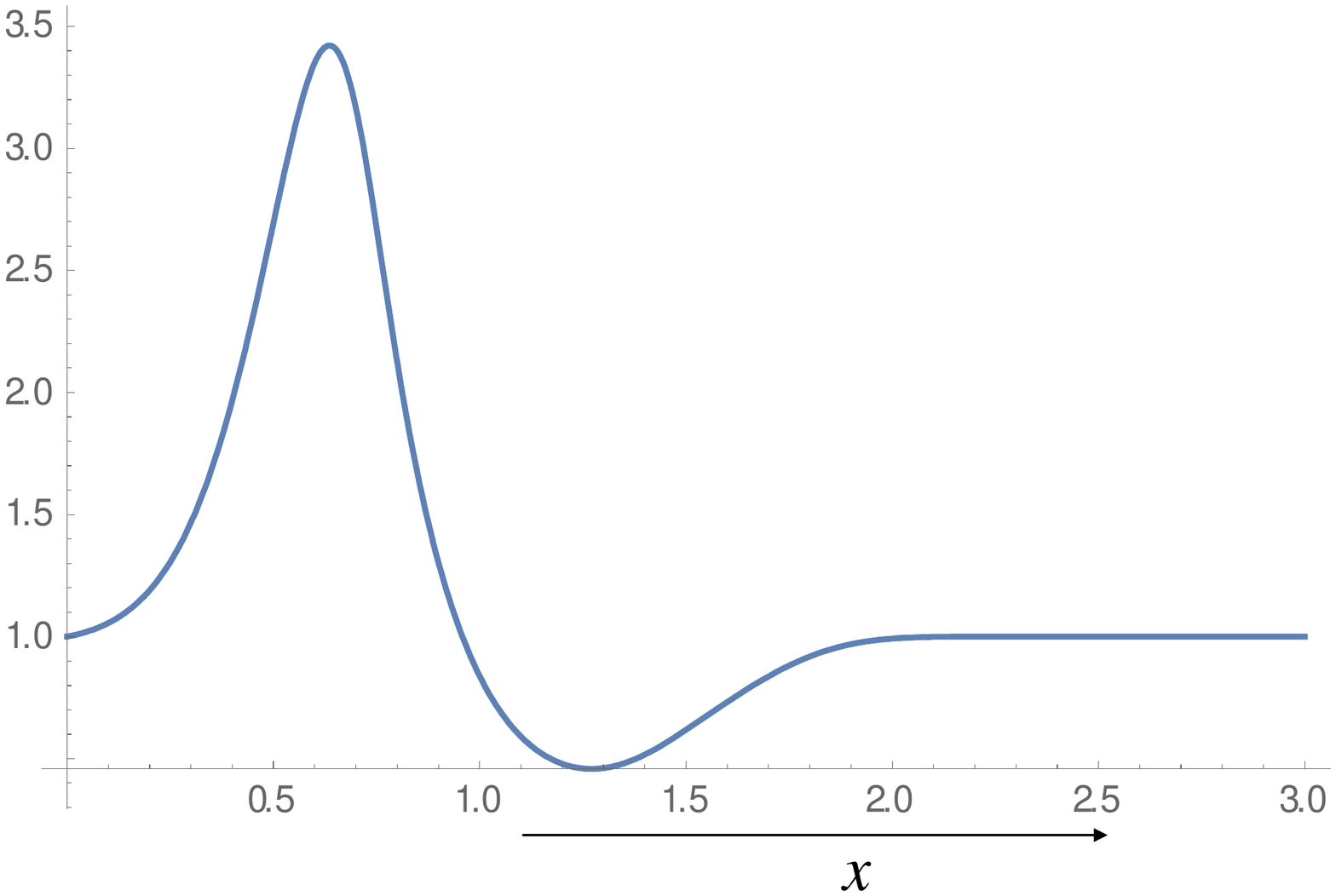}
\centering{(d) Counter-example \ref{ce3}(iv)}
\end{minipage}
\caption{\label{fig3}Graph of $F_{n:n}(x)/G_{n:n}(x)$} 
\end{figure}
\end{ce}
\section{Independent variables under random shocks}
Comparisons between maxima and minima are carried out in the last two sections on the assumption that each of the order statistics occur with certainty. Now, it is of interest to compare two samples stochastically when the order statistics of the two samples experience random shocks that may or may not lead to its failure. These models could arise natuarlly in reliability and actuarial sciences as described next.\\ 
Let us consider two samples of $n$ insurees each wherein the lifetimes of the insurees are independently distributed as GM. The objective then is to compare the maxima or minima of the two samples when each insuree of a sample receives random shocks that may lead to the death of the insuree. Let the random variable $U_{i}$ denote the lifetime of the $i$th insuree in a sample who experiences a random shock. Also, suppose for $i=1,2,\ldots, n$, $I_{i}$ denotes independent Bernoulli random variables, independent of $U_i$’s, with $E(I_i)=p_i$, which is referred to as the shock parameter. Then, the random shock impacts the $i$th insuree ($I_{i}= 1$) with probability $p_i$ or doesn't impact the $i$th insuree ($I_{i}= 0$) with probability $1-p_i$. Hence, the random variable $X_i =I_{i}T_{i}$ corresponds to the lifetime of the $i$th insuree under a shock. Under such a setting, in this section, we compare two largest (and smallest) order statistics from heterogeneous independent GM variables under random shocks through majorization. \\
For $i=1,2,\ldots,n$, let $U_i\sim GM\left(\alpha_i,\beta_i,\lambda_i\right)$ and $V_i\sim GM\left(\alpha_i^*,\beta_i^*,\lambda_i^*\right)$ be two sets of independent nonnegative random variables. Under random shocks, let us assume $X_i=U_iI_i$ and $Y_i=V_iI_i^*$. Thus, from (\ref{l3}) and for $x>0$, the survival functions of $X_i$ and $Y_i$ are given by 
\begin{equation}\label{5}
\overline {F_i}^X\left(x\right)=P(U_iI_i\geq x)=P(U_iI_i\geq x\mid I_i=1)P(I_i=1)=p_ie^{-\lambda_i x-\frac{\alpha_i}{\beta_i}\left(e^{\beta_i x}-1\right)}
\end{equation} 
and 
\begin{equation}\label{5a}
\overline{F_i}^Y\left(x\right)=P(V_iI_i^*\geq x)=P(V_iI_i^*\geq x\mid I_i^*=1)P(I_i^*=1)=p_i^* e^{-\lambda_i^* x-\frac{\alpha_i^*}{\beta_i^*}\left(e^{\beta_i^* x}-1\right)},
\end{equation}
respectively, where $E(I_i)=p_i$ and $E(I_i^*)=p_i^*$.
\subsection{\small Largest order statistics under random shocks}
Let $F^{X}_{n:n}\left(\cdot\right)$ and $F^Y_{n:n}\left(\cdot\right)$ be the cdf of $X_{n:n}$ and $Y_{n:n},$ respectively. Then, from (\ref{5}) and (\ref{5a}), we have, for $x>0$,
\begin{equation*}
F^{X}_{n:n}\left(x\right)=\prod_{i=1}^n\left[1- p_ie^{-\lambda_i x-\frac{\alpha_i}{\beta_i}\left(e^{\beta_i x}-1\right)}\right]
\end{equation*}
and
\begin{equation*}
F^{Y}_{n:n}\left(x\right)=\prod_{i=1}^n\left[1- p_i^* e^{-\lambda_i^* x-\frac{\alpha_i^*}{\beta_i^*}\left(e^{\beta_i^* x}-1\right)}\right],
\end{equation*}
with $F^{X}_{n:n}\left(0\right)=\prod_{i=1}^n(1- p_i)$ and $F^{Y}_{n:n}\left(0\right)=\prod_{i=1}^n(1-p_i^*)$.\\
Then, the following results show that under different restrictions on parameters, weakly submajorized shock parameter vector leads to larger life of the largest order statistic. 
\begin{t1}\label{th16}
For $i=1,2,\ldots, n$, let $U_i\sim GM\left(\alpha_i,\beta,\lambda_i\right)$ and $V_i\sim GM\left(\alpha_i,\beta,\lambda_i\right)$. Further, let $I_i~(I^{*}_i)$ be a set of independent Bernoulli random variables, independent of $U_i$'s ($V_i$'s) with $E(I_i)=p_i~(E(I^{*}_i)=p^{*}_i), i=1,2,...,n.$ Then, if $h:[0, 1]\rightarrow \mathbb{R}+$ is a differentiable, strictly increasing and convex function, 
 $$h(\mbox{\boldmath $p$})\succeq_w h(\mbox{\boldmath $p^{*}$)}\;\text{implies}\; X_{n:n}\ge_{st}Y_{n:n}$$ 
if $\mbox{\boldmath $\alpha$}, \mbox{\boldmath $\lambda$}\in \mathcal{D}_+ (\mathcal{E}_+)$ and $h(\mbox{\boldmath $p$)}\in \mathcal{E}_+ (\mathcal{D}_+)$, where $h(\mbox{\boldmath $p$})\equiv\left(h(p_1), h(p_2),\ldots, h(p_n)\right)$.
\end{t1}
{\bf Proof:} For fixed $\beta$ and $x\geq 0$, let
\begin{equation}\label{5.1}
 F^{X}_{n:n}\left(x\right)=\prod_{k=1}^n\left[1- h^{-1}(u_k)e^{-\lambda_k x-\frac{\alpha_k}{\beta}\left(e^{\beta x}-1\right)}\right]=\Psi_5\left(\mbox{\boldmath $u$}\right),
\end{equation}
where $h(p_i)=u_i$. Differentiating $\Psi_5\left(\mbox{\boldmath $u$}\right)$ with respect to $u_i$, we get 
\begin{equation*}
\frac{\partial \Psi_5}{\partial u_i}=-\prod_{k\neq i=1}^n{\left[1- h^{-1}(u_k)e^{-\lambda_k x-\frac{\alpha_k}{\beta}\left(e^{\beta x}-1\right)}\right]}\frac{dh^{-1}(u_i)}{du_i}e^{-\lambda_i x-\frac{\alpha_i}{\beta}\left(e^{\beta x}-1\right)}\leq 0,
\end{equation*}
showing that $\Psi_5\left(\mbox{\boldmath $u$}\right)$ is decreasing in each $u_i$. Thus, for all $x\geq 0$,
\begin{eqnarray}\label{5.2}
\frac{\partial \Psi_5}{\partial u_i}-\frac{\partial \Psi_5}{\partial u_j}&\stackrel{sign}{=}&\left[1- h^{-1}(u_i)e^{-\lambda_i x-\frac{\alpha_i}{\beta}\left(e^{\beta x}-1\right)}\right]\frac{dh^{-1}(u_j)}{du_j}e^{-\lambda_j x-\frac{\alpha_j}{\beta}\left(e^{\beta x}-1\right)}\nonumber\\
&-&\left[1- h^{-1}(u_j)e^{-\lambda_j x-\frac{\alpha_j}{\beta}\left(e^{\beta x}-1\right)}\right]\frac{dh^{-1}(u_i)}{du_i}e^{-\lambda_i x-\frac{\alpha_i}{\beta}\left(e^{\beta x}-1\right)}.
\end{eqnarray}
Now, for all $i\leq j$, $\alpha_i\geq (\leq)\alpha_j$, $\lambda_i\geq (\leq) \lambda_j$, $u_i\leq(\geq)u_j$ and $h(u)$ is increasing in $u$ gives 
\begin{equation}\label{5.3}
e^{-\lambda_i x-\frac{\alpha_i}{\beta}\left(e^{\beta x}-1\right)}\leq (\geq)e^{-\lambda_j x-\frac{\alpha_j}{\beta}\left(e^{\beta x}-1\right)}
\end{equation}
and $h^{-1}(u_i)\leq (\geq)h^{-1}(u_j)$, from which it can be shown that
\begin{equation}\label{5.4}
1-h^{-1}(u_i)e^{-\lambda_i x-\frac{\alpha_i}{\beta}\left(e^{\beta x}-1\right)}\geq (\leq)1-h^{-1}(u_j)e^{-\lambda_j x-\frac{\alpha_j}{\beta}\left(e^{\beta x}-1\right)}.
\end{equation}
Thus, from the fact that $h(u)$ is convex in $u$, and using equations (\ref{5.2})-(\ref{5.4}), it can be shown that
\begin{eqnarray*}
\frac{\partial \Psi_5}{\partial u_i}-\frac{\partial \Psi_5}{\partial u_j}&\stackrel{sign}{=}&\left[1- h^{-1}(u_i)e^{-\lambda_i x-\frac{\alpha_i}{\beta}\left(e^{\beta x}-1\right)}\right]\frac{dh^{-1}(u_j)}{du_j}e^{-\lambda_j x-\frac{\alpha_j}{\beta}\left(e^{\beta x}-1\right)}\nonumber\\
&-&\left[1- h^{-1}(u_j)e^{-\lambda_j x-\frac{\alpha_j}{\beta}\left(e^{\beta x}-1\right)}\right]\frac{dh^{-1}(u_i)}{du_i}e^{-\lambda_i x-\frac{\alpha_i}{\beta}\left(e^{\beta x}-1\right)}\\
&\geq (\leq)& 0.
\end{eqnarray*}
Thus, by Lemma \ref{l7} (Lemma \ref{l6}), $\Psi_5\left(\mbox{\boldmath $u$}\right)$ is s-concave in $\mbox{\boldmath $u$}$. So, as $\Psi_5\left(\mbox{\boldmath $u$}\right)$ is decreasing in each $u_i$, by Lemma \ref{l33}, the result gets proved.\hfill$\Box$\\
The following result can be proved along the same lines as Theorem \ref{th16} and so its proof is not presented.
\begin{t1}\label{th17}
For $i=1,2,\ldots, n$, let $U_i\sim GM\left(\alpha,\beta_i,\lambda_i\right)$ and $V_i\sim GM\left(\alpha,\beta_i,\lambda_i\right)$. Further, let $I_i~(I^{*}_i)$ be a set of independent Bernoulli random variables, independent of $U_i$'s ($V_i$'s) with $E(I_i)=p_i~(E(I^{*}_i)=p^{*}_i), i=1,2,...,n.$ Then, if $h:[0, 1]\rightarrow \mathbb{R}+$ is a differentiable, strictly increasing and convex function, then
 $$h(\mbox{\boldmath $p$})\succeq_w h(\mbox{\boldmath $p^{*}$)}\;\text{implies}\; X_{n:n}\ge_{st}Y_{n:n}$$ 
if $\mbox{\boldmath $\alpha$}, \mbox{\boldmath $\lambda$}\in \mathcal{D}_+ (\mathcal{E}_+)$ and $h(\mbox{\boldmath $p$)}\in \mathcal{E}_+ (\mathcal{D}_+)$, where $h(\mbox{\boldmath $p$})\equiv\left(h(p_1), h(p_2),\ldots, h(p_n)\right).$
\end{t1}
The following results show that there exists stochastic orderings between $X_{n:n}$ and $Y_{n:n}$ on conditions on different majorized parameters, when the shock parameter vector is the same for the two distributions.
\begin{t1}\label{th18}
For $i=1,2,\ldots, n$, let $U_i\sim GM\left(\alpha_i,\beta,\lambda_i\right)$ and $V_i\sim GM\left(\alpha_i^*,\beta,\lambda_i\right)$. Further, let $I_i$ be a set of independent Bernoulli random variables, independent of $U_i$'s ($V_i$'s) with $E(I_i)=p_i, i=1,2,...,n.$ Then, if $h:[0, 1]\rightarrow \mathbb{R}+$ is a differentiable, strictly increasing function,
 $$\mbox{\boldmath $\alpha$}\stackrel{w}{\succeq} \mbox{\boldmath $\alpha^{*}$}\;\text{implies}\; X_{n:n}\ge_{st}Y_{n:n}$$ 
if $\mbox{\boldmath $\alpha$}, \mbox{\boldmath $\alpha^*$}, \mbox{\boldmath $\lambda$}\in \mathcal{D}_+ (\mathcal{E}_+)$ and $h(\mbox{\boldmath $p$)}\in \mathcal{E}_+ (\mathcal{D}_+)$.
\end{t1}
{\bf Proof:} Consider
\begin{equation}\label{5.5}
 F^{X}_{n:n}\left(x\right)=\prod_{k=1}^n\left[1- h^{-1}(u_k)e^{-\lambda_k x-\frac{\alpha_k}{\beta}\left(e^{\beta x}-1\right)}\right]=\Psi_6\left(\mbox{\boldmath $\alpha$}\right),
\end{equation}
where $h(p_i)=u_i$. Then, 
\begin{equation*}
\frac{\partial \Psi_6}{\partial \alpha_i}=\prod_{k\neq i=1}^n{\left[1- h^{-1}(u_k)e^{-\lambda_k x-\frac{\alpha_k}{\beta}\left(e^{\beta x}-1\right)}\right]}h^{-1}(u_i)e^{-\lambda_i x-\frac{\alpha_i}{\beta}\left(e^{\beta x}-1\right)}\frac{e^{\beta x}-1}{\beta}\geq 0,
\end{equation*}
showing that $\Psi_6\left(\mbox{\boldmath $\alpha$}\right)$ is decreasing in each $\alpha_i$. So, for all $x\geq 0$, we have
\begin{eqnarray}\label{5.6}
\frac{\partial \Psi_6}{\partial \alpha_i}-\frac{\partial \Psi_6}{\partial \alpha_j}&\stackrel{sign}{=}&\left[1- h^{-1}(u_j)e^{-\lambda_j x-\frac{\alpha_j}{\beta}\left(e^{\beta x}-1\right)}\right]h^{-1}(u_i) e^{-\lambda_i x-\frac{\alpha_i}{\beta}\left(e^{\beta x}-1\right)}\nonumber\\
&-&\left[1- h^{-1}(u_i)e^{-\lambda_i x-\frac{\alpha_i}{\beta}\left(e^{\beta x}-1\right)}\right]h^{-1}(u_j) e^{-\lambda_j x-\frac{\alpha_j}{\beta}\left(e^{\beta x}-1\right)}\nonumber\\
&=& h^{-1}(u_i)e^{-\lambda_i x-\frac{\alpha_i}{\beta}\left(e^{\beta x}-1\right)}-h^{-1}(u_j)e^{-\lambda_j x-\frac{\alpha_j}{\beta}\left(e^{\beta x}-1\right)}.
\end{eqnarray}
Again, because $\mbox{\boldmath $\alpha$}, \mbox{\boldmath $\lambda$}\in \mathcal{D}_+ (\mathcal{E}_+)$ and $\mbox{\boldmath $u$}\in \mathcal{E}_+ (\mathcal{D}_+)$ and $h(u)$ is increasing in $u$, then for all $i\leq j$, $\alpha_i\geq (\leq)\alpha_j$, $\lambda_i\geq (\leq) \lambda_j$, $u_i\leq(\geq)u_j$ gives 
\begin{equation*}
h^{-1}(u_i)e^{-\lambda_i x-\frac{\alpha_i}{\beta}\left(e^{\beta x}-1\right)}\leq (\geq)1-h^{-1}(u_j)e^{-\lambda_j x-\frac{\alpha_j}{\beta}\left(e^{\beta x}-1\right)},
\end{equation*}
which, by (\ref{5.6}), yields $\frac{\partial \Psi_6}{\partial \alpha_i}-\frac{\partial \Psi_6}{\partial \alpha_j}\leq (\geq) 0.$
Thus, by Lemmas \ref{l6} (Lemma \ref{l7}) and \ref{l33} the result is proved.\hfill$\Box$\\
By following the same lines as Theorem \ref{th18}, the following results can be established, and so their proofs are not presented.
\begin{t1}\label{th19}
For $i=1,2,\ldots, n$, let $U_i\sim GM\left(\alpha_i,\beta,\lambda_i\right)$ and $V_i\sim GM\left(\alpha_i,\beta,\lambda_i^*\right)$. Further, let $I_i$ be a set of independent Bernoulli random variables, independent of $U_i$'s ($V_i$'s) with $E(I_i)=p_i, i=1,2,...,n.$ Then, if $h:[0, 1]\rightarrow \mathbb{R}+$ is a differentiable, strictly increasing function, 
 $$\mbox{\boldmath $\lambda$}\stackrel{w}{\succeq} \mbox{\boldmath $\lambda^{*}$}\;\text{implies}\; X_{n:n}\ge_{st}Y_{n:n}$$ 
if $\mbox{\boldmath $\alpha$}, \mbox{\boldmath $\lambda$}, \mbox{\boldmath $\lambda^*$}\in \mathcal{D}_+ (\mathcal{E}_+)$ and $h(\mbox{\boldmath $p$)}\in \mathcal{E}_+ (\mathcal{D}_+)$.
\end{t1}
\begin{t1}\label{th20}
For $i=1,2,\ldots, n$, let $U_i\sim GM\left(\alpha,\beta_i,\lambda_i\right)$ and $V_i\sim GM\left(\alpha,\beta_i,\lambda_i^*\right)$. Further, let $I_i$ be a set of independent Bernoulli random variables, independent of $U_i$'s ($V_i$'s) with $E(I_i)=p_i, i=1,2,...,n.$ Then, if $h:[0, 1]\rightarrow \mathbb{R}+$ is a differentiable, strictly increasing function,
 $$\mbox{\boldmath $\lambda$}\stackrel{w}{\succeq} \mbox{\boldmath $\lambda^{*}$}\;\text{implies}\; X_{n:n}\ge_{st}Y_{n:n}$$ 
if $\mbox{\boldmath $\alpha$}, \mbox{\boldmath $\lambda$}, \mbox{\boldmath $\lambda^*$}\in \mathcal{D}_+ (\mathcal{E}_+)$ and $h(\mbox{\boldmath $p$)}\in \mathcal{E}_+ (\mathcal{D}_+)$.
\end{t1}
For fixed $\alpha$, the result below shows that weak submajorization ordering between $\frac{1}{\mbox{\boldmath $\beta$}}$ and $\frac{1}{\mbox{\boldmath $\beta^*$}}$ implies stochastic ordering between $X_{n:n}$ and $Y_{n:n}$.
\begin{t1}\label{th21}
For $i=1,2,\ldots, n$, let $U_i\sim GM\left(\alpha,\beta_i,\lambda_i\right)$ and $V_i\sim GM\left(\alpha,\beta_i^*,\lambda_i\right)$. Further, let $I_i$ be a set of independent Bernoulli random variables, independent of $U_i$'s ($V_i$'s) with $E(I_i)=p_i, i=1,2,...,n.$ Then, if $h:[0, 1]\rightarrow \mathbb{R}+$ is a differentiable, strictly increasing function, 
 $$\frac{1}{\mbox{\boldmath $\beta$}}\succeq_w \frac{1}{\mbox{\boldmath $\beta^{*}$}}\;\text{implies}\; X_{n:n}\ge_{st}Y_{n:n}$$ 
if $\mbox{\boldmath $\beta$}, \mbox{\boldmath $\beta^*$}, \mbox{\boldmath $\lambda$}\in \mathcal{D}_+ (\mathcal{E}_+)$ and $h(\mbox{\boldmath $p$)}\in \mathcal{E}_+ (\mathcal{D}_+)$.
\end{t1}
{\bf Proof:} Consider
\begin{equation*}
 F^{X}_{n:n}\left(x\right)=\prod_{k=1}^n\left[1- h^{-1}(u_k)e^{-\lambda_k x-\alpha z_k}\right]=\Psi_7\left(\mbox{\boldmath $z$}\right),
\end{equation*}
where $z_i=\frac{e^{\beta_i x}-1}{\beta_i}$ and $\mbox{\boldmath $z$}=\left(z_1,z_2,\ldots,z_n\right)$. Differentiating $\Psi_7\left(\mbox{\boldmath $z$}\right)$ with respect to $z_i,$ we obtain
\begin{equation*}
\frac{\partial \Psi_7}{\partial z_i}=\prod_{k\neq i=1}^n{\left[1- h^{-1}(u_k)e^{-\lambda_k x-\alpha z_k}\right]}h^{-1}(u_i)\alpha e^{-\lambda_i x-\alpha z_i}\geq 0,
\end{equation*}
yielding that $\Psi_7\left(\mbox{\boldmath $z$}\right)$ is increasing in each $z_i$. Thus, for $i\leq j$, we obtain
\begin{eqnarray}\label{5.7}
\frac{\partial \Psi_7}{\partial z_i}-\frac{\partial \Psi_7}{\partial z_j}&\stackrel{sign}{=}&\left[1- h^{-1}(u_j)e^{-\lambda_j x-\alpha z_j}\right]h^{-1}(u_i) e^{-\lambda_i x-\alpha z_i}-\left[1- h^{-1}(u_i)e^{-\lambda_i x-\alpha z_i}\right]h^{-1}(u_j) e^{-\lambda_j x-\alpha z_j}\nonumber\\
&=& h^{-1}(u_i)e^{-\lambda_i x-\alpha z_i}-h^{-1}(u_j)e^{-\lambda_j x-\alpha z_j} \leq (\geq) 0,
\end{eqnarray}
for $\mbox{\boldmath $\beta$}, \mbox{\boldmath $\beta^*$}, \mbox{\boldmath $\lambda$}\in \mathcal{D}_+ (\mathcal{E}_+)$, $\mbox{\boldmath $u$}\in \mathcal{E}_+ (\mathcal{D}_+)$ and $h(u)$ is increasing function in $u$. Thus, by Lemmas \ref{l6} (Lemma \ref{l7}) and \ref{l33}, it can be proved that $\mbox{\boldmath $z$}\stackrel{w}{\succeq} \mbox{\boldmath $z^{*}$}\;\text{implies}\; X_{n:n}\ge_{st}Y_{n:n}.$ Now taking $\beta_i=-\frac{1}{\nu_i}$ and proceeding as in Theorem \ref{th15}, the result can be proved.\hfill$\Box$
\subsection{\small Smallest order statistics under random shocks}
Let $\overline F^{X}_{1:n}\left(\cdot\right)$ and $\overline F^Y_{1:n}\left(\cdot\right)$ be the survival functions of $X_{1:n}$ and $Y_{1:n},$ respectively. Then, from (\ref{5}) and (\ref{5a}), we have, for $x>0$,
\begin{equation}\label{5.8}
\overline F^{X}_{1:n}\left(x\right)=\prod_{i=1}^n {p_i}e^{-\sum_{i=1}^n\left[\lambda_i x+\frac{\alpha_i}{\beta_i}\left(e^{\beta_i x}-1\right)\right]}
\end{equation}
and
\begin{equation}\label{5.9}
\overline F^{Y}_{1:n}\left(x\right)=\prod_{i=1}^n {p_i^*}e^{-\sum_{i=1}^n\left[\lambda_i^* x+\frac{\alpha_i^*}{\beta_i^*}\left(e^{\beta_i^* x}-1\right)\right]},
\end{equation}
with $\overline F^{X}_{1:n}\left(0\right)=\prod_{i=1}^n p_i$ and $\overline F^{Y}_{1:n}\left(0\right)=\prod_{i=1}^np_i^*$.\\
The following results show that under certain restrictions on parameters, hazard rate ordering exists between $X_{1:n}$ and $Y_{1:n}$. While the proof of Theorem \ref{th22} is presented, Theorems \ref{th23} and \ref{th24} can be proved in a similar manner and so their proofs are not presented for conciseness.
\begin{t1}\label{th22}
For $i=1,2,\ldots, n$, let $U_i\sim GM\left(\alpha_i,\beta_i,\lambda_i\right)$ and $V_i\sim GM\left(\alpha_i^*,\beta_i,\lambda_i\right)$. Further, let $I_i(I_i^*)$ be a set of independent Bernoulli random variables, independent of $U_i$'s ($V_i$'s) with $E(I_i)=p_i (E(I_i^*)=p_i^*), i=1,2,...,n.$ Then, if $\mbox{\boldmath $\alpha$}, \mbox{\boldmath $\alpha^*$}\in \mathcal{D}_+ (\mathcal{E}_+)$, $\mbox{\boldmath $\beta$}\in \mathcal{D}_+$ and $\prod_{i=1}^n p_i\geq (\leq)\prod_{i=1}^n p_{i}^*$, 
 $$\mbox{\boldmath $\alpha$}\stackrel{m}{\succeq} \mbox{\boldmath $\alpha^{*}$}\;\text{implies}\; X_{1:n}\le_{hr}(\ge_{hr})Y_{1:n}.$$ 
\end{t1}
{\bf Proof:} From equations \ref{11} and \ref{12}, using \ref{5.8} and \ref{5.9}, we can write, for $x>0,$ 
$$\frac{\overline{F}^{X}_{1:n}\left(x\right)}{\overline{F}^{Y}_{1:n}\left(x\right)}=\frac{\prod_{i=1}^n p_i}{\prod_{i=1}^n p_i^{*}}\frac{e^{-\sum_{i=1}^n\left[\lambda_i x+\frac{\alpha_i}{\beta_i}\left(e^{\beta_i x}-1\right)\right]}}{e^{-\sum_{i=1}^n\left[\lambda_i x+\frac{\alpha_i^*}{\beta_i}\left(e^{\beta_i x}-1\right)\right]}}=\frac{\prod_{i=1}^n p_i}{\prod_{i=1}^n p_i^{*}}\frac{\overline{G}_{1:n}\left(x\right)}{\overline{H}_{1:n}\left(x\right)},$$ 
which is increasing (decreasing) in $x,$ by Theorem~\ref{th4}.\\ 
Now, using the fact that $\lim_{x\rightarrow 0-}\frac{\overline{F}^{X}_{1:n}\left(x\right)}{\overline{F}^{Y}_{1:n}\left(x\right)}=1,$ we have $$1\leq(\geq)\frac{\prod_{i=1}^n p_i}{\prod_{i=1}^n p_i^{*}}\Rightarrow \lim_{x\rightarrow 0-}\frac{\overline{F}^{X}_{1:n}\left(x\right)}{\overline{F}^{Y}_{1:n}\left(x\right)}\leq(\geq) \frac{\prod_{i=1}^n p_i}{\prod_{i=1}^n p_i^{*}}\frac{\overline{G}_{1:n}\left(0\right)}{\overline{H}_{1:n}\left(0\right)}=\frac{\overline{F}^{X}_{1:n}\left(0\right)}{\overline{F}^{Y}_{1:n}\left(0\right)},$$ proving that $\frac{\overline{F}^{X}_{1:n}\left(x\right)}{\overline{F}^{Y}_{1:n}\left(x\right)}$ is increasing (decreasing) at $x=0.$ This proves the result.  \hfill$\Box$\\
\begin{t1}\label{th23}
For $i=1,2,\ldots, n$, let $U_i\sim GM\left(\alpha_i,\beta_i,\lambda_i\right)$ and $V_i\sim GM\left(\alpha_i,\beta_i^*,\lambda_i\right)$. Further, let $I_i(I_i^*)$ be a set of independent Bernoulli random variables, independent of $U_i$'s ($V_i$'s) with $E(I_i)=p_i (E(I_i^*)=p_i^*), i=1,2,...,n.$ Then, if $\mbox{\boldmath $\alpha$}, \mbox{\boldmath $\beta$}, \mbox{\boldmath $\beta^*$}\in \mathcal{D}_+ (\mathcal{E}_+)$ and $\prod_{i=1}^n p_i\geq (\leq)\prod_{i=1}^n p_{i}^*$,
 $$\mbox{\boldmath $\beta$}\stackrel{m}{\succeq} \mbox{\boldmath $\beta^{*}$}\;\text{implies}\; X_{1:n}\le_{hr} Y_{1:n}.$$ 
\end{t1}
\begin{t1}\label{th24}
For $i=1,2,\ldots, n$, let $U_i\sim GM\left(\alpha_i,\beta_i,\lambda_i\right)$ and $V_i\sim GM\left(\alpha_i,\beta_i,\lambda_i^*\right)$. Further, let $I_i(I_i^*)$ be a set of independent Bernoulli random variables, independent of $U_i$'s ($V_i$'s) with $E(I_i)=p_i (E(I_i^*)=p_i^*), i=1,2,...,n.$ Then, if $\mbox{\boldmath $\alpha$}, \mbox{\boldmath $\beta$}, \mbox{\boldmath $\beta^*$}\in \mathcal{D}_+ (\mathcal{E}_+)$ and $\prod_{i=1}^n p_i\geq (\leq)\prod_{i=1}^n p_{i}^*$,
 $$\sum_{k=1}^n \lambda_k\geq \sum_{k=1}^n \lambda_k^*\;\text{implies}\; X_{1:n}\le_{hr} Y_{1:n}.$$ 
\end{t1}

\end{document}